\documentclass[useAMS,usenatbib]{mn2e}

\usepackage{graphicx}
\usepackage{amssymb}
\usepackage{epstopdf}
\usepackage{graphics}
\usepackage{epsfig}
\usepackage{color}
\usepackage{amsmath}

\title[Pulsar Probes of Black Hole Gravity]{Fast Spinning Pulsars as Probes of Massive Black Holes' Gravity}
\author[Dinesh Singh, Kinwah Wu and Gordon E. Sarty]{Dinesh Singh$^{1,2}$\thanks{E-mail:
dinesh.singh@uregina.ca (DS); kw@mssl.ucl.ac.uk (KW); gordon.sarty@usask.ca (GES)}, Kinwah Wu$^{3}$ and Gordon E.~Sarty$^{2,4}$ \\
\\
$^{1}$ Department of Physics, University of Regina, Regina, SK, S4S~0A2, Canada  \\
$^{2}$ Department of Physics and Engineering Physics, University of Saskatchewan, Saskatoon, SK, S7N~5E3, Canada  \\
$^{3}$ Mullard Space Science Laboratory, University College London, Holmbury St Mary, Surrey RH5 6NT    \\
$^{4}$ Department of Psychology, University of Saskatchewan, Saskatoon, SK, S7N~5E3, Canada
}

\begin{document}
\onecolumn

\date{Submitted 2013}

\pagerange{\pageref{firstpage}--\pageref{lastpage}} \pubyear{2012}

\maketitle

\label{firstpage}

\begin{abstract}
Dwarf galaxies and globular clusters may contain intermediate mass black holes ($10^{3}$ to $10^{5}{\rm M}_{\odot}$) in their cores. Estimates of $\sim 10^{3}$ neutron stars in the central parsec of the Galaxy and similar numbers in small elliptical galaxies and globular clusters along with an estimated high probability of ms-pulsar formation in those environments has led many workers to propose the use of ms-pulsar timing to measure the mass and spin of intermediate mass black holes. Models of pulsar motion around a rotating black hole generally assume geodesic motion of a ``test'' particle in the Kerr metric. These approaches account for well-known effects like de Sitter precession and the Lense-Thirring effect but they do not account for the non-linear effect of the pulsar's stress-energy tensor on the space-time metric. Here we model the motion of a pulsar near a black hole with the Mathisson-Papapetrou-Dixon (MPD) equations. Numerical integration of the MPD equations for black holes of mass $2 \times 10^{6}$, $10^{5}$ and $10^{3}$ M$_{\odot}$ shows that the pulsar will not remain in an orbital plane with motion vertical to the plane being largest relative to the orbit's radial dimensions for the lower mass black holes. The pulsar's out of plane motion will lead to timing variations that are up to $\sim 10 \mu$s different from those predicted by planar orbit models. Such variations might be detectable in long term observations of millisecond pulsars. If pulsar signals are used to measure the mass and spin of intermediate mass black holes on the basis of dynamical models of the received pulsar signal then the out of plane motion of the pulsar should be part of that model.
\end{abstract}


\section{Introduction}
\label{sec-intro}

The presence of astrophysical black holes is inferred from various observations,
 such as the powerful electromagnetic radiation emitted by distant quasars.
Although we have not yet ``seen'' black holes directly,
   it will soon be possible to image the massive central black hole
   in the Galaxy \citep[see][]{Doeleman08} and some nearby galaxies,
   e.g.\  M87, using submm VLBI observations \citep[see][]{Doeleman12,Dexter12,Asada12}.
At present the strongest evidence for a massive black hole at the centre of the Galaxy comes from monitoring the motions of stars in the Sgr A* region.
These observations have established that a large amount of unseen mass,
  $\approx 4.2 \times 10^{6}\, {\rm M}_{\odot}$  \citep{Ghez08, Gillessen09},
  is enclosed within a volume having a radius $< 0.01$~pc at the Galactic Centre
  \citep[see][]{Eckart97, Ghez98}.
The simple explanation for this unseen mass is a massive black hole
  \citep[see][]{Schoedel02, Ghez08, Gillessen09}.
Naturally we would ask if this massive black hole is rapidly rotating or it is slowly rotating.
Knowing the black hole's rotational rate has important astrophysical implications.
It indicates how the black hole has evolved and perhaps how it was formed --
 whether the black hole was built up by the merging of smaller black holes or simply by accreting a large amount of gas.

A rotating black hole drags the space-time around it so stars and gas respond differently to Kerr and Schwarzschild black holes.
X-ray spectroscopy of relativistic lines has been used to determine the spin of several black holes in active galactic nuclei
  \citep[e.g.\ for MCG-60-30-15,][]{Iwasawa96}.
Theoretical calculations  \citep[e.g.][]{Laor91, Kojima91} show
  that the profiles of relativistic emission lines, such as the Fe~K$\alpha$ line
  emitted from the surface of the accretion disk around a black hole, depend on the black hole's spin. However,
  in practice the reliability of the relativistic line method of black-hole spin measurement
  depends also on how well we model the accretion flow
  and on how well we understand the radiative processes that give rise to the diagnostic lines
  in the disk region close to the black hole event horizon \citep[see e.g.][]{Fuerst07, Svoboda09, Younsi12}.
It is always a challenging task to measure the spin of a black hole much less to measure it with accuracy,
  be it a stellar-mass black hole in a binary system or a supermassive black hole in an active galactic nucleus.
As shown in theoretical calculations,
  the parameter space is actually degenerate for the relativistic X-ray line profiles
  \citep{Laor91, Kojima91, Fuerst04},
  thus one needs to resolve this issue properly to obtain a reliable black hole spin measurement.
As for the black hole in the Galactic Centre, the
  lack of X-ray activity \citep{Baganoff03} in the current epoch
  implies an absence of an opaque gas accretion disk on which the relativistic X-ray lines are expected to form.
Alternative methods for determining the black hole spin are therefore much needed.

Observations have shown correlations between the mass of central black holes
  and the properties of the bulges of their host galaxies.
In particular, a relatively tight $M$-$\sigma$ correlation
  is found for the nearby big galaxies \citep{Ferrarese00, Gebhardt00},
  where $M$ is the mass of the central black hole and $\sigma$ is the velocity dispersion of the stars in the galactic bulge.
For the Galaxy,
   the mass estimate of the central black hole
   and the measured velocity dispersion of the stars in the Galactic bulge
   are consistent with the empirical $M$-$\sigma$ relation derived for external galaxies \citep[see][]{Gueltekin09}.
The most massive astrophysical black holes known to date have masses around $\sim 1.5 \times 10^{10}\, {\rm M}_{\odot}$ 
  \citep[e.g. the central black hole in NGC~1277,][]{vandenBosch12}.
Nuclear black holes with masses below $10^{6}\, {\rm M}_{\odot}$ in galaxies
  are not firmly established by stellar dynamics or by reverberation mapping \citep{Peterson04},
but there are observations indicating that some Seyfert galaxies may contain nuclear black holes
  in the mass range of $10^{5}-10^{6}\, {\rm M}_{\odot}$ \citep{Greene07, Xiao11}.
The inclusion of small-bulge (low-mass) galaxies appears to steepen the slope of the $M$-$\sigma$ relation \citep{Graham11}.
It is still unclear whether the least massive dwarf galaxies contain a central black hole
  (with $M_{\rm bh} \sim 10^{4}\, {\rm M}_{\odot}$) similar to the big elliptical galaxies.
The lower mass limit for the central black holes in galaxies is not certain.

Further extrapolation of the $M$-$\sigma$ relation to low-mass stellar spheroids
  implies that globular clusters would have nuclear black holes with mass $\sim 10^{3 }-10^{4}\, {\rm M}_{\odot}$ 
  \citep[see][]{Lutzgendorf13}.
There have been active searches for the intermediate-mass black holes
  (IMBH, black holes with masses $\sim 10^{2}-10^{4}{\rm M}_{\odot}$) in globular clusters as well as in dwarf galaxies.
While there are claims of the discovery of IMBHs in globular clusters,
  there are also counterclaims of non-detection \citep[see e.g.\ the discussions in][]{vanderMarel10}.
It is of great importance to accurately measure black hole masses in low-mass stellar spheroids
 and to properly resolve the issues regarding the low-end of the mass spectrum of non-stellar black holes.

Here, in this work, we analyze the orbital motion of millisecond pulsars (ms-pulsars, fast spinning neutron stars)
   around a rotating black hole taking into account the effect of the pulsar's stress-energy tensor on the Kerr metric of the black hole.
The compactness of neutron stars and the large mass ratios between nuclear black holes and the neutron stars
  allow a point-particle approximation for the neutron star,
  without compromising a proper treatment of the interaction
  between the spin of the neutron star  and the black hole spin as manifested by the interaction
  between the spin of the neutron star and the the curvature of space-time induced by the black hole's gravity.
Thus, the dynamics of these systems are well described by the Mathisson-Papapetrou-Dixon (MPD) equations
  for spinning test particles in an external space-time.
We show how the orbital dynamics of a ms-pulsar is determined by spin-curvature coupling
  when it revolves around a black hole
  and how the orbital dynamics depend on the spin as well as the mass ratio between the black hole and the pulsar.
In particular, we show that motion of the pulsar out of the usual orbital plane is substantial, relative to the orbital extent, if the mass of the rotating black hole is low enough.
We organize the paper as follows.
In \S 2 we present the formulation of the dynamics of systems containing a spinning neutron star orbiting around a massive black hole.
In \S3 we give the scheme for solving the MPD equations
  and solutions of some example systems with parameters of astronomical interest.
The significance of such binary systems and resulting astrophysical implications are discussed in \S4.
Throughout in this work, unless otherwise stated, we use the natural unit system with $c =G =1$,
  where $c$ is the speed of light, and $G$ is the gravitational constant.
We also adopt a signature of $+2$ for the space-time metric tensor.

\section{Spin interaction between a fast spinning neutron star and a black hole}
\label{sec-MPD}

Consider a pulsar, a spinning neutron star, orbiting around a massive black hole.
As the black hole is much more massive than a neutron star, i.e.\ $M_{\rm bh} \gg M_{\rm ns}$,
  the pulsar can be treated as a test mass.
The pulsar's motion is then determined by a background gravitational field provided by the black hole
  and its dynamical interaction with this field.

Let the mass of the pulsar (neutron star) be $m\ ( = M_{\rm ns}$) and the mass of the black hole be $M\ (=M_{\rm bh})$.
The neutron star has a radius $R_{\rm ns}$,
  which is much smaller than the Schwarzschild radius of the black hole, $R_{\rm s}(M_{\rm bh})$,
  and the orbital separation between the centre-of-mass of the two objects is $r$.
The 4-velocity of the centre of mass of the pulsar is represented by
\begin{eqnarray}
 u^{\mu} & = &  \frac{dx^{\mu}}{d \tau} \ ,
\end{eqnarray}
   where $\tau$ is the proper time along its world line.
The equation of motion of the pulsar is given by
\begin{eqnarray}
  {T^{\mu \nu}}_{; \nu} &  = & 0 \ ,
\end{eqnarray}
  where $T^{\mu \nu}$  is the energy-momentum tensor.
The tensor can be expanded into an infinite set of multipole moments \citep{Dixon74}.
The first two moments are the momentum vector $p^{\nu}$ and the spin tensor $s^{\mu \nu}$.
Their corresponding equations of motion are
\begin{eqnarray}
   \frac{Dp^{\mu}}{d\tau}  \  & = & -\frac{1}{2} {R^{\mu}}_{\nu\alpha\beta}u^{\nu} s^{\alpha \beta} +{\cal F}^{\mu} \ ;  \label{eqn-MPD1} \\
   \frac{Ds^{\mu\nu}}{d\tau}  & = & p^{\mu}u^{\nu} - p^{\nu}u^{\mu} +{\cal T}^{\mu \nu} \    \label{eqn-MPD2}
\end{eqnarray}
\citep{Mathisson37, Papapetrou51, Dixon74}.
The Dixon force ${\cal F}^{\mu}$ and torque ${\cal T}^{\mu \nu}$
  are determined by the quadrupole and the higher momentum of the pulsar, when it has a non-zero finite size.
A supplementary condition to the equations of motion (\ref{eqn-MPD1}) and (\ref{eqn-MPD2}) is required
  for a proper specification of trajectory of the pulsar's centre of mass and this is taken as
\begin{eqnarray}
  s^{\mu \nu} p_{\nu} & = & 0 \ .  \label{eqn-MPD3}
\end{eqnarray}
Note that other choices for the supplementary condition can also be made in order to fully determine the equations of motion.
The set of equations (\ref{eqn-MPD1}) and  (\ref{eqn-MPD2}) with the supplementary condition (\ref{eqn-MPD3})
   are known as the Mathisson-Papapetrou-Dixon (MPD) equations  \citep[see][]{Mashhoon06}.

For $m\ll M < r$ and $R_{\rm ns} \ll r$, the interaction is dominated by the lowest order moments.
As an approximation, we may ignore the quadrupole and higher-order moments
  and set ${\cal F}^{\mu}=0$ and ${\cal T}^{\mu \nu}=0$ in the MPD equations,
  resulting in the reduced MPD equations.
This leads to the usual expression for the mass of the pulsar
\begin{eqnarray}
  m & = &  \sqrt{-p^{\mu}p_{\mu}} \ ,
\end{eqnarray}
  which is a constant of motion, as shown by taking the covariant time derivative
  of (\ref{eqn-MPD3}) and contracting with $D p_{\mu}/d\tau$.
The spin vector of the pulsar is given by
\begin{eqnarray}
  s_{\mu} & = &  -\frac{1}{2m} \epsilon_{\mu \nu \alpha \beta} p^{\nu} s^{\alpha \beta}  \ .
\end{eqnarray}
 where the Levi-Civita tensor $\epsilon_{\mu \nu \alpha \beta} = \sqrt{-g} ~\sigma_{\mu \nu \alpha \beta}$,
 with the permutation $\sigma_{0123} =1$.
The spin tensor can be expressed in terms of the spin vector
\begin{eqnarray}
 s^{\mu \nu} & = &  \frac{1}{m}  \epsilon^{\mu \nu \alpha \beta}p_{\alpha} s_{\beta} \ .
\end{eqnarray}
It follows from the reduced MPD equations that
\begin{eqnarray}
  s^{2} & = &   s^{\mu} s_{\mu} = \frac{1}{2} s^{\mu\nu} s_{\mu\nu} \ ,
\end{eqnarray}
   which is also a constant of motion, as shown from using (\ref{eqn-MPD3}).

As $m \ll M < r$, we have $R_{\rm M}= s/m \ll r$,
  where $R_{\rm M}$ is the M{\o}ller radius of the pulsar.
This implies that the dipole-dipole interaction is significantly weaker than the pole-dipole interaction,
  thus allowing us to apply the approximation scheme of \cite{Chicone05}.
The scheme is based on the condition that
\begin{eqnarray}
 \left( \frac{p^{\mu}}{m} - u^{\mu} \right) & \sim & \frac{M}{r}\left[\frac{s}{mr} \right]^{2} \ll 1 \  .
\end{eqnarray}
To the first order in $s/(mr)$, $p^{\nu} \approx m u^{\mu}$, i.e.\ the momentum and velocity 4-vectors are parallel to each other.
In a more intuitive sense, this corresponds to the situation
  that the kinetic energy of the pulsar is insignificant in comparison to the rest mass energy.
With this approximation, the reduced MPD equations become
\begin{eqnarray}
   \frac{Du^{\mu}}{d\tau}  \  & = & -\frac{1}{2m} {R^{\mu}}_{\nu\alpha\beta}u^{\nu} s^{\alpha \beta}   \ ;  \label{eqn-mMPD1} \\
   \frac{Ds^{\mu\nu}}{d\tau}  & \approx & 0  \  ;   \label{eqn-mMPD2}  \\
  s_{\mu \nu} u^{\nu} & \approx  & 0 \     \label{eqn-mMPD3}
\end{eqnarray}
\citep{Chicone05, Mashhoon06}.
In the case of a slowly spinning pulsar,
  the above equations are reduced to the geodesic equation for the orbital motion
  of a point-like spinless object in an external gravitational field due to a black hole \citep[see][]{Ehlers04}.

\section{Orbital dynamics of the spinning neutron star}
\label{sec-orb}

The solution to the MPD equations has been derived for various settings
  \citep[e.g.][]{Semerak99,Bini04,Singh05,Mashhoon06,Kyrian07,Singh08,Plyatsko11}.
\cite{Mashhoon06} investigated various solution schemes for Kerr black holes
  and found that the approximation schemes may not always capture all the essential aspects of spin multi-pole interactions
  for general situations (such as those of astrophysical interest).
In this work we consider a full numerical approach and integrate the reduced MPD equations directly.
For our calculations the reduced MPD equations take the following form:
\begin{eqnarray}
 \frac{dp^{\alpha}}{d\tau} & = & -\Gamma^{\alpha}_{\mu\nu}p^{\mu}u^{\nu}
    + \lambda \left(\frac{1}{2m} {R^{\alpha}}_{\beta \rho \sigma} {\epsilon^{\rho \sigma}}_{\mu\nu} s^{\mu}p^{\nu}u^{\beta}   \right)  \ ;
       \label{MPD-x1}   \\
 \frac{ds^{\alpha}}{d\tau} & = &  -\Gamma^{\alpha}_{\mu\nu}s^{\mu}u^{\nu}
    + \lambda \left(\frac{1}{2m^{3}} R_{\gamma \beta \rho \sigma}{\epsilon^{\rho \sigma}}_{\mu\nu}  s^{\mu}p^{\nu}s^{\gamma }u^{\beta}   \right)p^{\alpha} \ ;
     \label{MPD-x2}  \\
 \frac{dx^{\alpha}}{d\tau} & = & u^{\alpha} =   - \frac{p^{\delta}u_{\delta}}{m^{2}}\left(     p^{\alpha}
     + \frac{1}{2}  \frac{\lambda \left(s^{\alpha \beta} R_{\beta \gamma \mu \nu} p^{\gamma} s^{\mu \nu} \right)}
      {m^{2}  + \lambda \left(  R_{\mu \nu \rho \sigma}s^{\mu \nu}s^{\beta\sigma}/4  \right ) }    \right)
     \label{MPD-x3}
\end{eqnarray}
\citep{Singh05, Mashhoon06}
with $\tau$ as the affine parameter.
Although $\tau$ has the freedom to not be  the proper time,
  here we choose $\tau$ as the proper time
   such that $g_{\mu\nu} u^{\mu} u^{\nu}=-1$ throughout the motion of the orbiting pulsar.
A dimensionless parameter $\lambda$ is introduced in the above equations
  to tag the terms for MPD spin-curvature coupling, as in \cite{Singh05}.
For $\lambda = 1$, the contribution of spin-curvature coupling to the evolution of the pulsar spin and the orbital dynamics is included;
  for $\lambda = 0$ the contribution of spin-curvature coupling is omitted and hence the evolution of the pulsar spin is strictly due to parallel transport.

The space-time around a rotating black hole is given by the Kerr metric, which is
\begin{eqnarray}
- d\tau^2 & = & - \biggl( 1- {{2Mr} \over \Sigma}\biggr)dt^2
         - {{4aMr \sin^2\theta} \over \Sigma}dtd\phi
        + {\Sigma \over \Delta}dr^2
  + \Sigma d\theta^2
    + \biggl( r^2+a^2 + {{2a^2Mr \sin^2\theta} \over \Sigma} \biggr)
                  \sin^2\theta d\phi^2
\end{eqnarray}
  in Boyer-Lindquist coordinates,
  where $\Sigma = r^2+a^2\cos^2\theta$, $\Delta = r^2 -2Mr +a^2$,
  and the three vector $(r,\theta,\phi$) corresponds to (pseudo-)spherical polar coordinates.
The parameter $a/M$ specifies the spin of the black hole,
  with $a/M = 0$ corresponding to the Schwarzschild black hole
  and $a/M = 1$ to the maximally rotating Kerr black hole.
  

\begin{table}
\begin{center}
\caption{Eccentricities of the orbits. The numbers in brackets denote values that vary from orbit to orbit. 
\label{tab-1}} 
\begin{tabular}{lcccccc}
\hline
Black Hole Spin & \multicolumn{3}{c}{$\xleftarrow{\hspace*{5em}} a/M = 0.1 \xrightarrow{\hspace*{5em}}$} &  \multicolumn{3}{c}{$\xleftarrow{\hspace*{5em}} a/M =  -0.1  \xrightarrow{\hspace*{5em}}$}\\
Black Hole Mass & $10^{3}$ M$_{\odot}$ & $10^{5}$ M$_{\odot}$ & $2 \times 10^{6}$ M$_{\odot}$ & $10^{3}$ M$_{\odot}$ & $10^{5}$ M$_{\odot}$ & $2 \times 10^{6}$ M$_{\odot}$ \\
\hline
{$e$ for $r = 10 M$} & 0.1946(5) & 0.1945(9)  & 0.1945(9)  & 0.2094(1)  & 0.2093(6)  &  0.2093(6) \\
{$e$ for $r = 20 M$} & 0.10662(6)  & 0.10661(1)  &  0.10661(1) & 0.1085(5)  & 0.10854(4)  & 0.10854(3)  \\
{$e$ for $r = 30 M$} & 0.08173659(3) &  0.0817289(7) & 0.08172890(5)  & 0.08245(2)  &  0.082445(6) &  0.082445(4) \\
{$e$ for $r = 40 M$}  & 0.068784(3)  & 0.068779(4)  & 0.06877(9)  & 0.069155(3)  & 0.069150(3)  &  0.06915(0) \\
\hline \hline
Black Hole Spin & \multicolumn{3}{c}{$\xleftarrow{\hspace*{5em}} a/M = 0.5 \xrightarrow{\hspace*{5em}}$} &  \multicolumn{3}{c}{$\xleftarrow{\hspace*{5em}} a/M =  -0.5 \xrightarrow{\hspace*{5em}}$}\\
Black Hole Mass & $10^{3}$ M$_{\odot}$ & $10^{5}$ M$_{\odot}$ & $2 \times 10^{6}$ M$_{\odot}$ & $10^{3}$ M$_{\odot}$ & $10^{5}$ M$_{\odot}$ & $2 \times 10^{6}$ M$_{\odot}$ \\
\hline
{$e$ for $r = 10 M$} & 0.172(0) &  0.1719(7) & 0.1719(7)  & 0.2483(8) & 0.2483(3)  & 0.2483(2)  \\
{$e$ for $r = 20 M$} & 0.10319(8) & 0.10318(5)  & 0.10318(5)  & 0.11292(6)  & 0.11290(7)  & 0.11290(7)  \\
{$e$ for $r = 30 M$} & 0.08041(5) & 0.08040(8)  & 0.08040(8)  & 0.08400(9)  & 0.084000(5)  & 0.084000(5)  \\
{$e$ for $r = 40 M$} & 0.068088(7)  & 0.068084(3)  & 0.068084(2)  & 0.06994(6)  & 0.069940(5)  & 0.069940(5)  \\
\hline \hline
Black Hole Spin & \multicolumn{3}{c}{$\xleftarrow{\hspace*{5em}} a/M = 0.99 \xrightarrow{\hspace*{5em}}$} &  \multicolumn{3}{c}{$\xleftarrow{\hspace*{5em}} a/M =  -0.99 \xrightarrow{\hspace*{5em}}$}\\
Black Hole Mass & $10^{3}$ M$_{\odot}$ & $10^{5}$ M$_{\odot}$ & $2 \times 10^{6}$ M$_{\odot}$ & $10^{3}$ M$_{\odot}$ & $10^{5}$ M$_{\odot}$ & $2 \times 10^{6}$ M$_{\odot}$ \\
\hline
{$e$ for $r = 10 M$} & 0.1531(8) & 0.1531(5)  & 0.1531(5)  & 0.3195(1)  & 0.3194(3)  & 0.3194(3)  \\
{$e$ for $r = 20 M$} & 0.09969(4)  & 0.09968(3)  &  0.09968(3) & 0.11933(5)  & 0.11931(7)  & 0.11931(8)  \\
{$e$ for $r = 30 M$} & 0.07898(7) & 0.07898(1)  & 0.07898(1)  & 0.086156(5)  & 0.08614(6)  & 0.08614(6)  \\
{$e$ for $r = 40 M$} & 0.06731(7)   & 0.06731(3)  &  0.06731(2) & 0.071008(6)  & 0.071002(6)  & 0.071002(6)  \\
\hline
\end{tabular}
\end{center}
\end{table} 


In this work,
we consider three kinds of astrophysical black holes with masses:
(i) $M = 2.0 \times 10^{6}{\rm M}_{\odot}$,
     which is about the same as the mass of the black hole in the Galactic Centre;
(ii) $M = 10^{5}{\rm M}_{\odot}$,
     which is at the low end of the empirical $M$-$\sigma$ relation for black holes in galactic bulges
     and is similar to those of the lower-mass black holes of the Seyferts in the study of  \cite{Greene07};
 and (iii) $M=10^{3}{\rm M}_{\odot}$,
    which is the mass of the smaller expected intermediate-mass black holes  in the globular clusters
    obtained by extrapolating the $M$-$\sigma$ relation.
The pulsar has a mass $m=1.5~\!{\rm M}_{\odot}$,
  and a spin period $P_{\rm s} = 1~\!{\rm ms}$.
Its initial orbital radius $r$ has values ranging from $10M$ to $40M$.
Assuming prograde motion with respect to the spin of the black hole, the initial
orbital angular motion chosen for all cases is $J = J_{\rm circ} + \Delta J$, where
\begin{eqnarray}
J_{\rm circ} & = & \frac{\left[1 + \left(\frac{a}{M}\right)\left(\frac{M}{r}\right)^2 - \frac{a}{M} \left(\frac{M}{r}\right)^{3/2} \right] M}
{\sqrt{\frac{M}{r}\left[1 - \frac{3M}{r} + \frac{2a}{M}\left(\frac{M}{r}\right)^{3/2}\right]}}
\label{J-circ}
\end{eqnarray}
    \citep{Raine05} is the orbital angular momentum for strictly circular motion,
   with $\Delta J = 0.2 \, M$ to generate precessing quasi-elliptical orbits beginning at periastron.
The eccentricity of an orbit is defined as $e = (r_{\rm a}-r_{\rm p})/(r_{\rm a}+r_{\rm p})$,
    where $r_{\rm a}$ is the radius of the orbit at apastron and $r_{\rm p}$ is the radius of the orbit at periastron.
 For Keplerian orbits and orbits of test particles in a Schwarzschild metric, the eccentricity is constant.
 For the orbits modelled here, the eccentricity varies slightly from orbit to orbit.
 The values of eccentricity for the cases considered here are given in Table~\ref{tab-1}.
 
Orbits of a pulsar around a slowly rotating ($|a/M|=0.1$) black hole with $M = 2.0\times 10^{6}{\rm M}_{\odot}$
  at various initial distances from the black hole's centre had the following characteristics.
At $r=40M$ the orbit is a precessing ellipse and the deviation from being elliptical can be regarded
   as a perturbation caused by the relativistic orbital (de Sitter) precession. 
This result is not too surprising. 
For a sufficiently large distance (i.e.\ $r \gg M$), relativistic effects are not very prominent,
  and the spin-pole coupling between the pulsar and black hole and the spin-orbit coupling are weak.
In that case we expect that the orbital motion of the pulsar would be similar to that of a planar Keplerian orbit in a Newtonian space-time.
As the distance between the pulsar and the black hole decreases, the orbit will further deviate from a simply precessing elliptical orbit.
The orbit begins to exhibit complexities at $r =20M$.
For smaller $r$ the orbital motion can no longer be considered Keplerian in any approximate sense.
At $r = 10M$, relativistic spin coupling is clearly an important factor in determining the pulsar's orbital dynamics.
For slowly rotating black holes there are only subtle differences between the prograde and the retrograde orbits.
The orbits of a pulsar around a fast rotating ($|a/M| =0.99$) black hole are more complex.
At large distances ($r>40M$), the orbits are very similar to those of the slowing rotating black hole.
The differences between the fast rotating and the slowly rotating black holes become more obvious at smaller distances.
At $r=10M$ the differences between the prograde and the retrograde orbits around a fast spinning black holes
  are easily distinguishable,
  with the retrograde orbits showing complex patterns resembling that of the precession of elliptical orbits.
Moreover, the difference between the prograde orbits around a slowing rotating and fast rotating black hole is also noticeable in terms of motion out of the $x$--$y$ plane. The motion of the pulsar in the $\lambda=0$ case, in which the spin-curvature coupling is not modeled, remains in the $x$--$y$ plane where it is similar to the $x$--$y$ motion computed with $\lambda = 1$.

Comparison between orbits of pulsars around a $10^{5}{\rm M}_{\odot}$ black hole with $r = 10M$,
  for black-hole spins $|a/M|= 0.1$, $0.5$ and $0.99$ revealed the following effects.
The general trend is that the complexity of the orbit increases with the black-hole spin rate,
  and the increase is more for the retrograde orbit than the prograde orbit.
The orbits of the pulsars around a $2.0\times 10^{6}{\rm M}_{\odot}$ black hole and a $10^{3}{\rm M}_{\odot}$ black hole
  show only very slight differences from the $10^{5}{\rm M}_{\odot}$ case.
  

\begin{figure}
\begin{center}
  \includegraphics[width=0.325\textwidth]{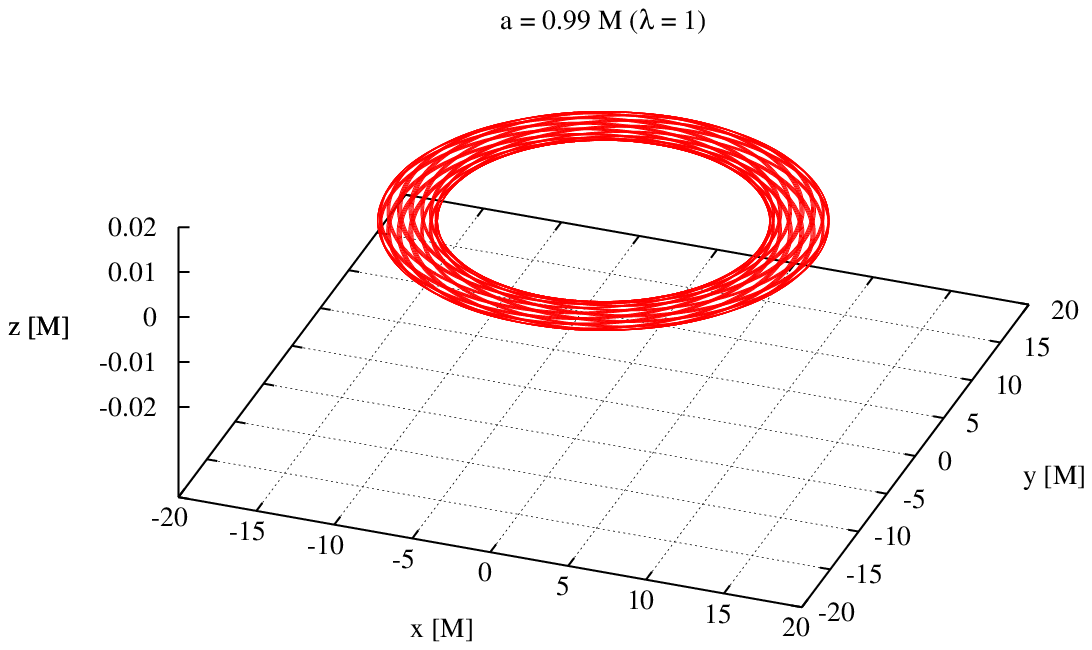}
  \includegraphics[width=0.325\textwidth]{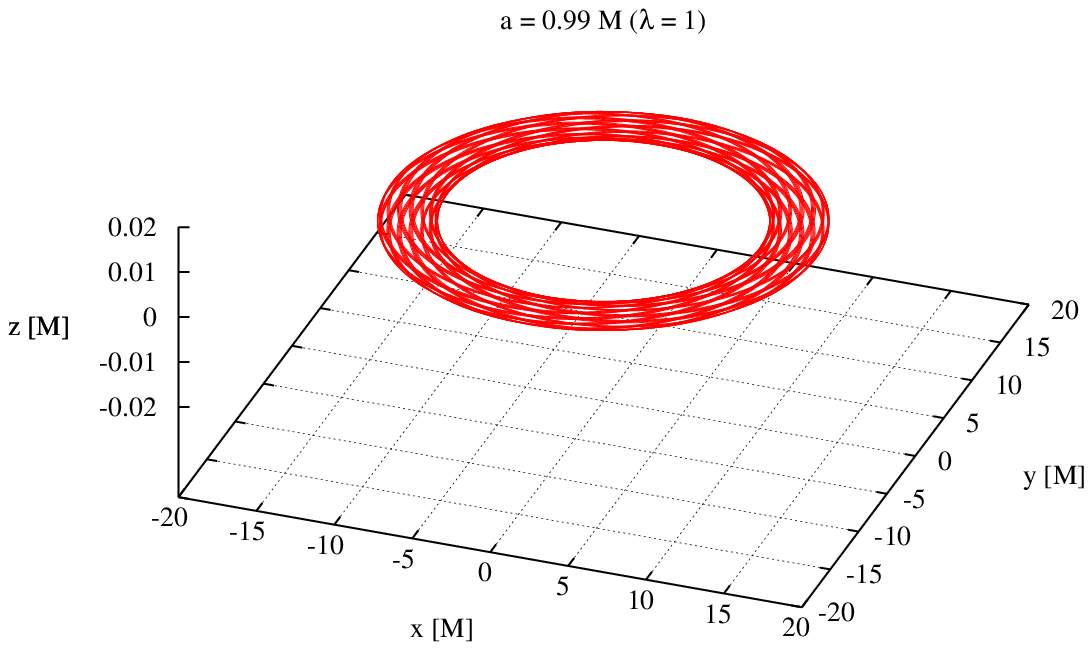}
  \includegraphics[width=0.325\textwidth]{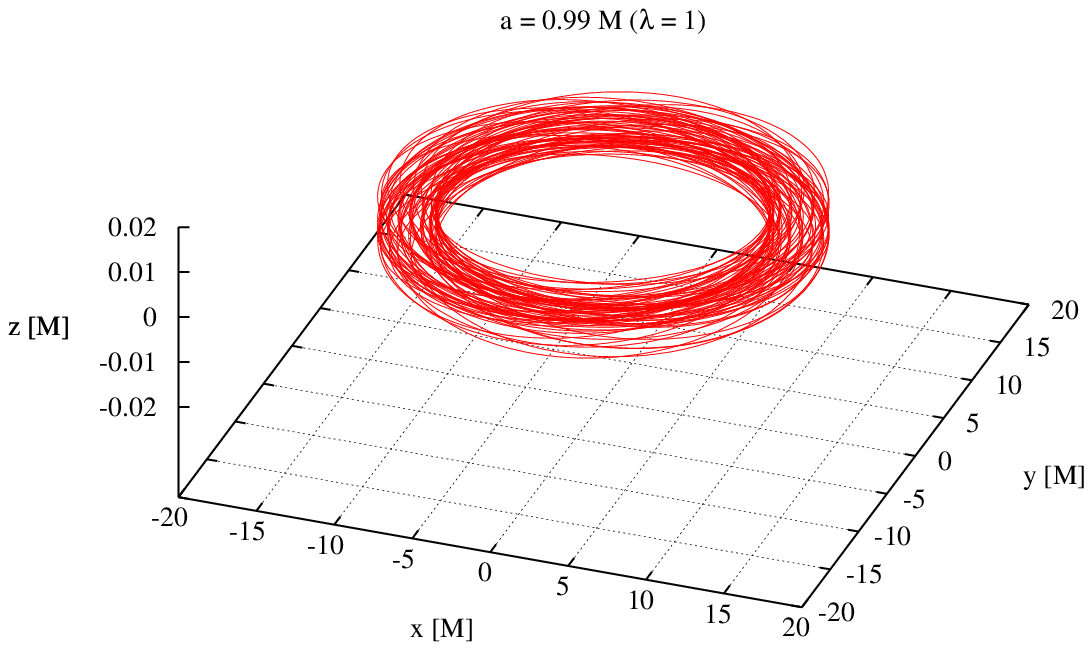}
  \includegraphics[width=0.325\textwidth]{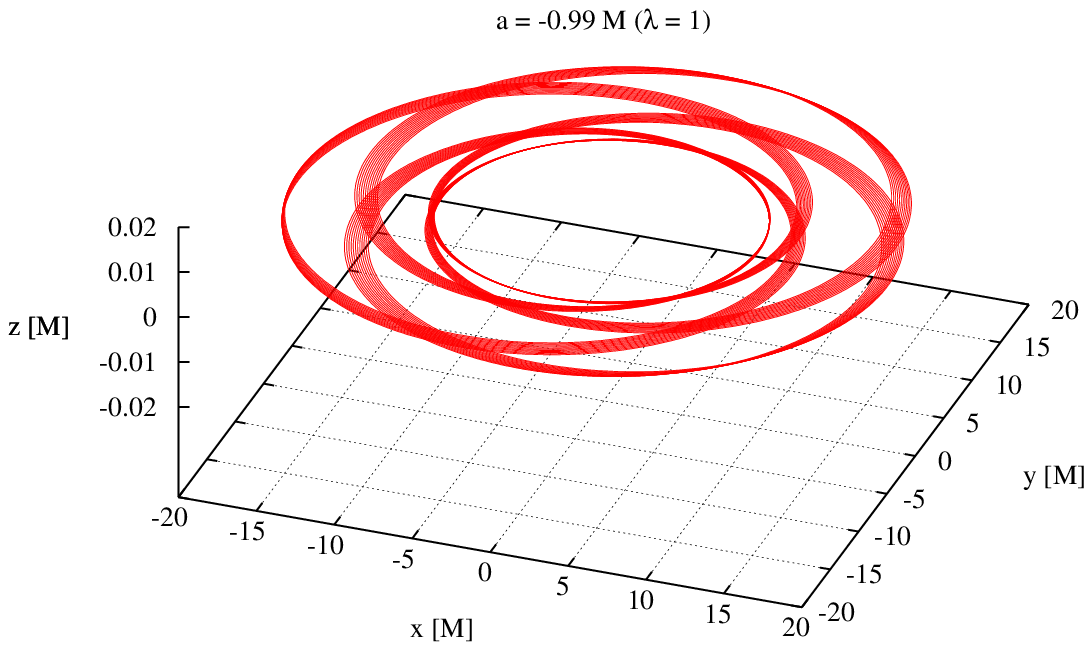}
  \includegraphics[width=0.325\textwidth]{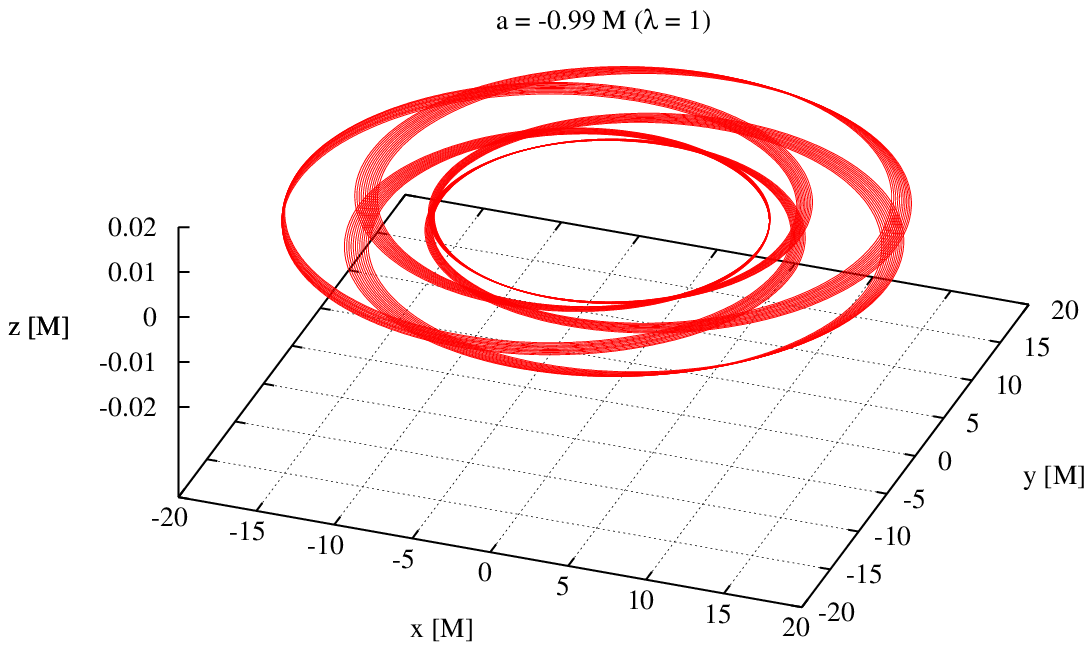}
  \includegraphics[width=0.325\textwidth]{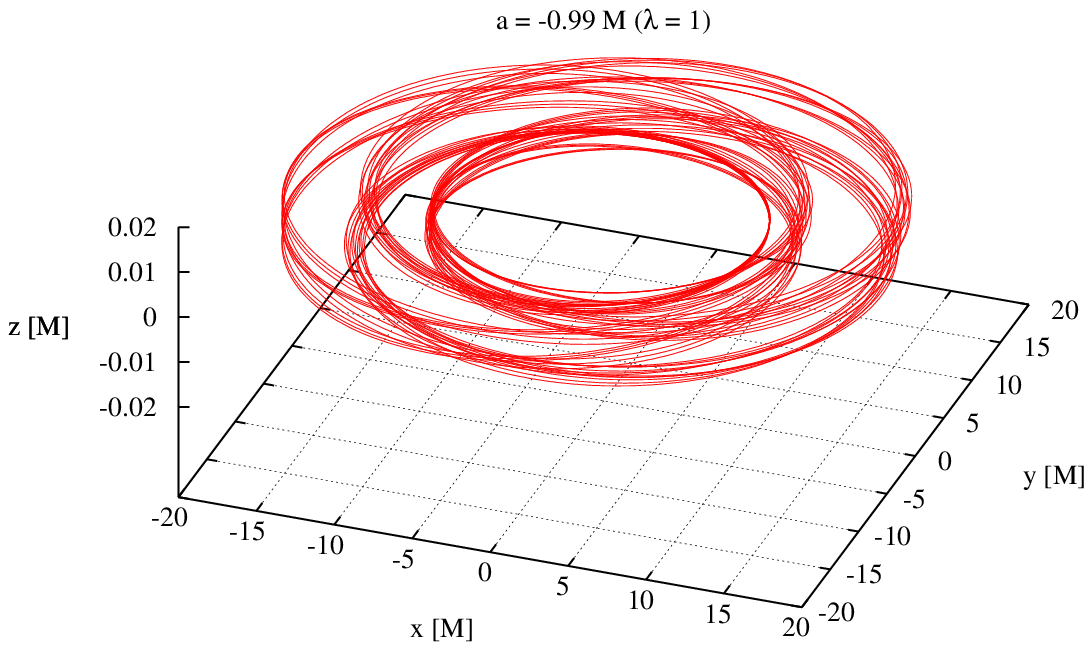}
\vspace*{0.25cm}
\caption{
  The orbit of a pulsar around a black hole for $M=  2.0 \times 10^{6}{\rm M}_{\odot}$ (left panels)
     $10^{5}{\rm M}_{\odot}$ (middle panels) and $10^{3}{\rm M}_{\odot}$ (right panels).
  The black-hole spin parameter $a/M = \pm0.99$ and the initial orbital separation $r=10M$.
  The prograde orbits are shown in the top row and the retrograde orbits in the bottom row.
  Positive $a$ corresponds to a pulsar in a prograde orbit with respect to the black hole's spin,
    and negative $a$ corresponds to a pulsar in a retrograde orbit.
 The centre of the black hole is located at $(0,0,0)$.
  The orbital normal vector of the pulsar and the spin vector of the black hole are
   in parallel (for a prograde pulsar orbit) or in anti-parallel (for a retrograde pulsar orbit) initially.
 The spin vector of the pulsar is initially oriented at a tilting angle of $\pi/4$ towards the black hole.
 The dimensionless parameter $\lambda$ is set to 1. In all cases the pulsar does not stay in an orbital plane, the $z$ motion being most obvious in the lower mass black hole cases in these plots. The out-of-plane motion is due to the $\lambda$-dependent terms in the MPD equations (\ref{MPD-x1}), (\ref{MPD-x2}) and (\ref{MPD-x3}). 
 In calculations with $\lambda = 0$, the orbital motion is the same in the $x$--$y$ plane as the cases shown here but the pulsar does not move out of the plane.
  }
\label{fig-5}
\end{center}
\end{figure}


The orbital dynamics of the pulsar depends on the mass ratio between the central black hole and the pulsar, $M/m$. The orbits are more complex for smaller $M/m$. Even in moderate conditions, such as in systems with an orbital separation $r = 30 M$ and a black-hole spin $|a/M|=0.1$, we can still distinguish the orbit for the pulsars around the black holes with mass  $10^{3}{\rm M}_{\odot}$ from those of $M= 2.0\times 10^{6}{\rm M}_{\odot}$ and  $10^{5}{\rm M}_{\odot}$. For the more extreme conditions, such as for systems with an orbital separation $r = 10 M$ and a black-hole spin $|a/M|=0.99$ (Figure \ref{fig-5}), the orbits clearly show a strong dependence on the mass ratio $M/m$, but the dependence is relative as discussed next. 

\begin{figure}
\begin{center}
\vspace*{-5em}
  \includegraphics[width=0.325\textwidth]{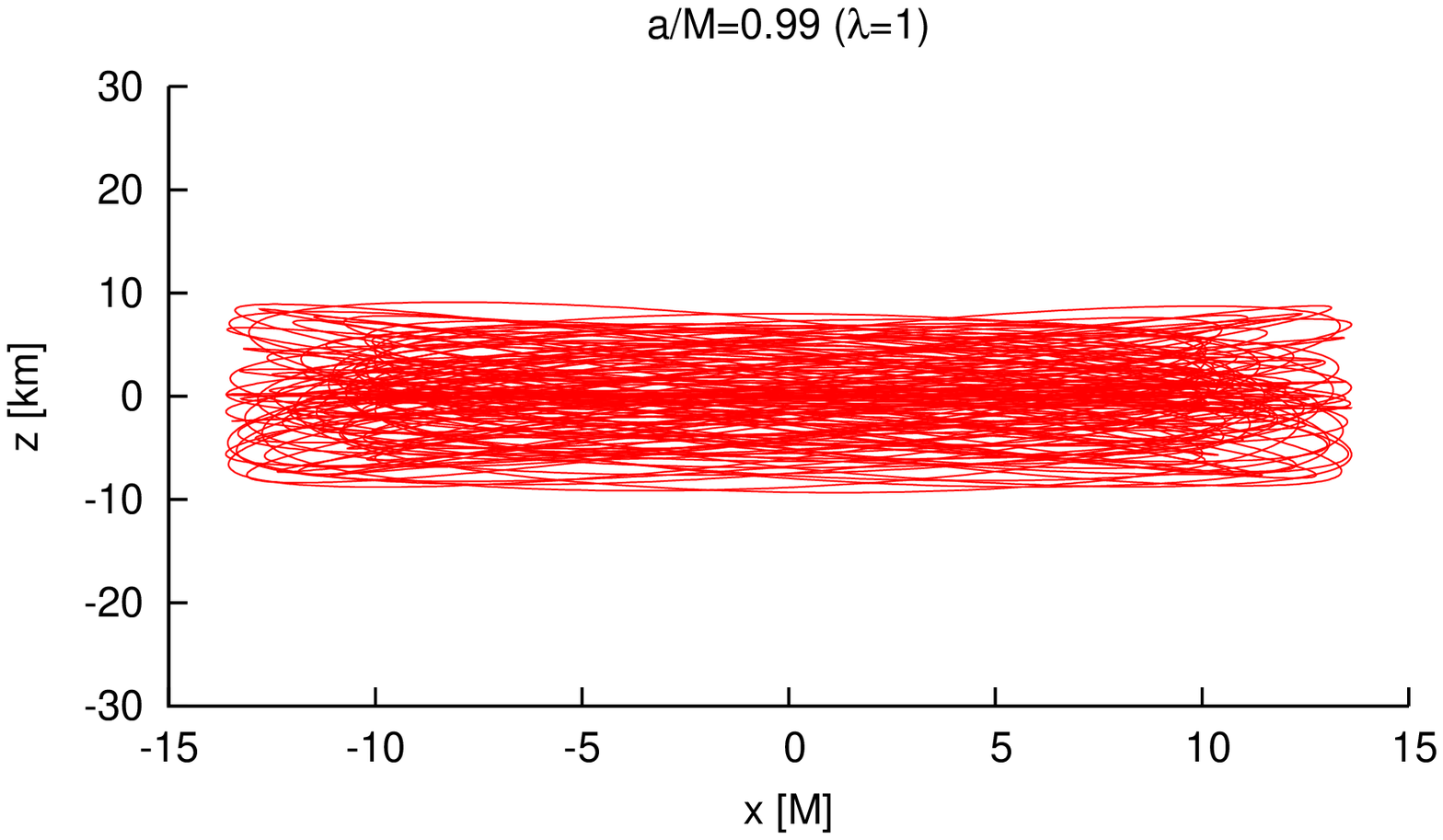}
  \includegraphics[width=0.325\textwidth]{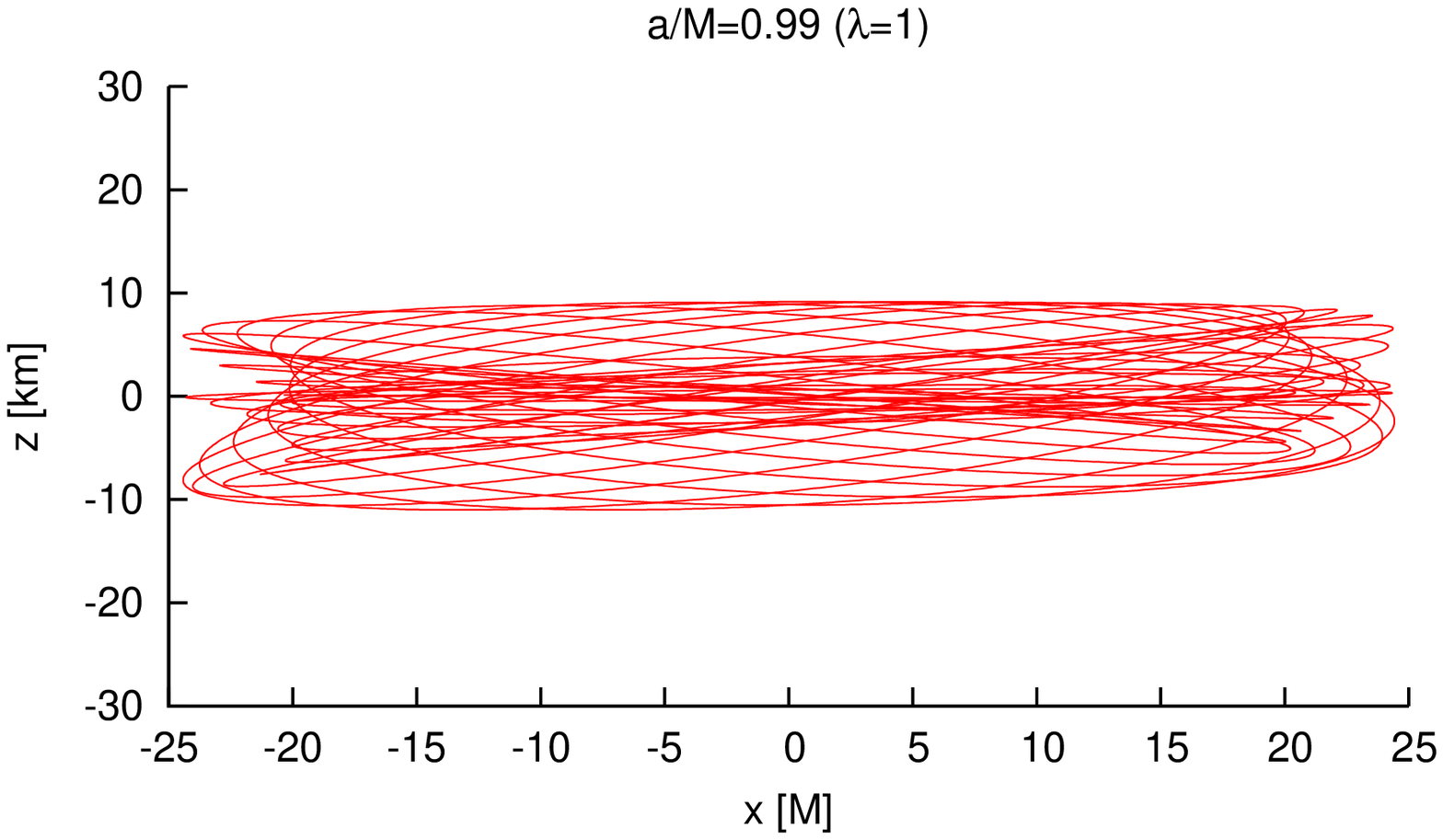}
  \includegraphics[width=0.325\textwidth]{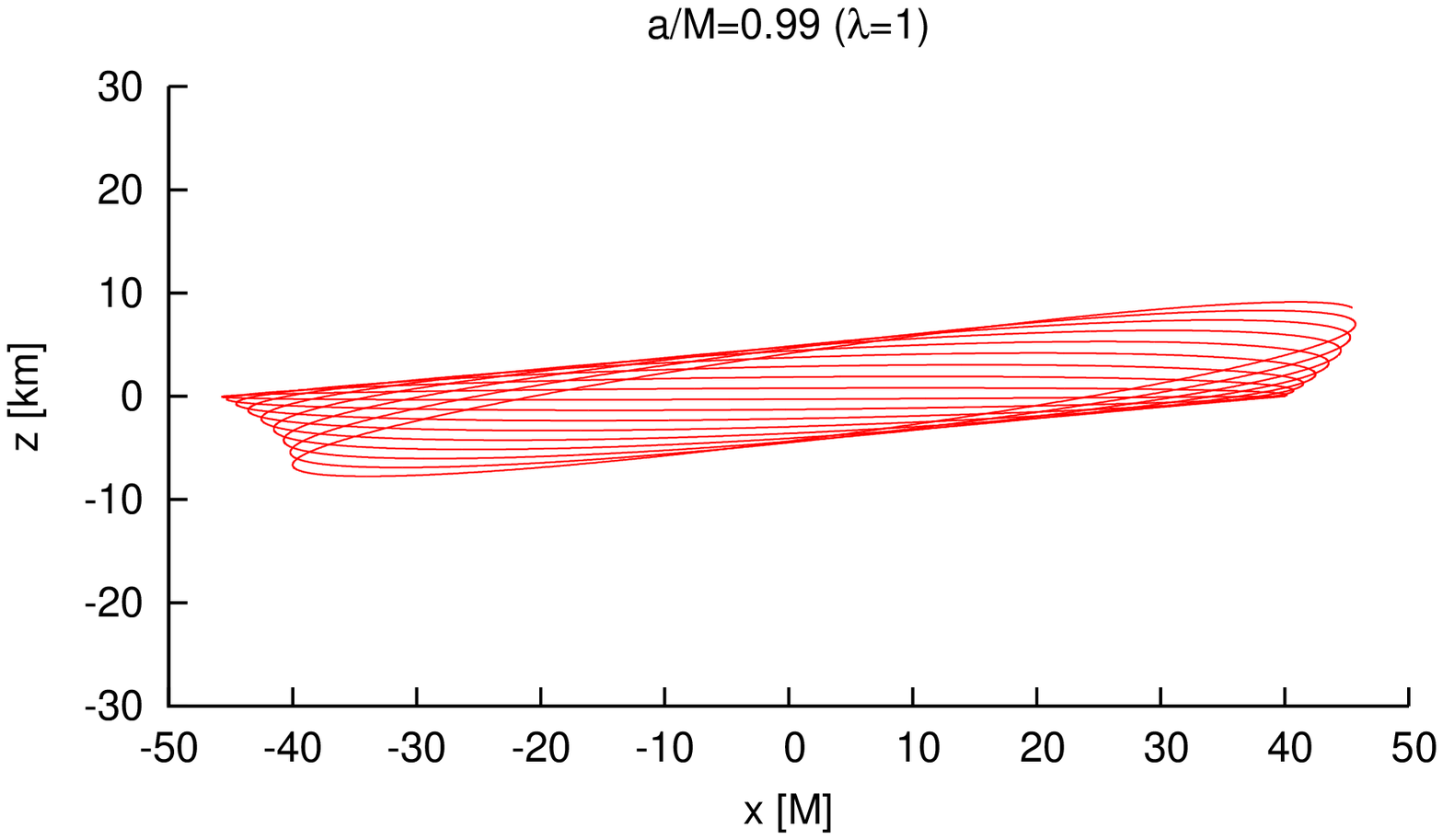}

\vspace*{-14em}

  \includegraphics[width=0.325\textwidth]{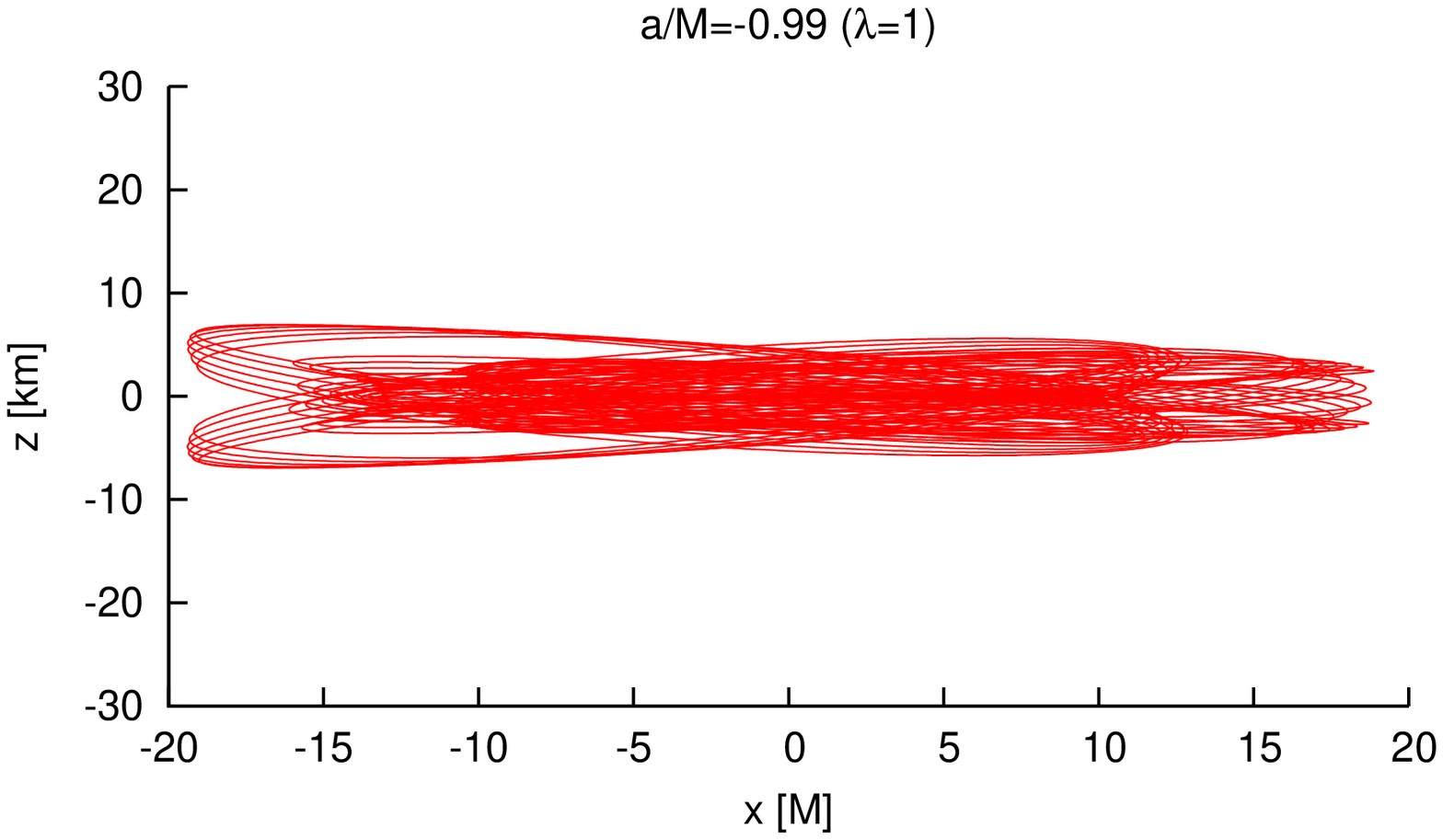}
  \includegraphics[width=0.325\textwidth]{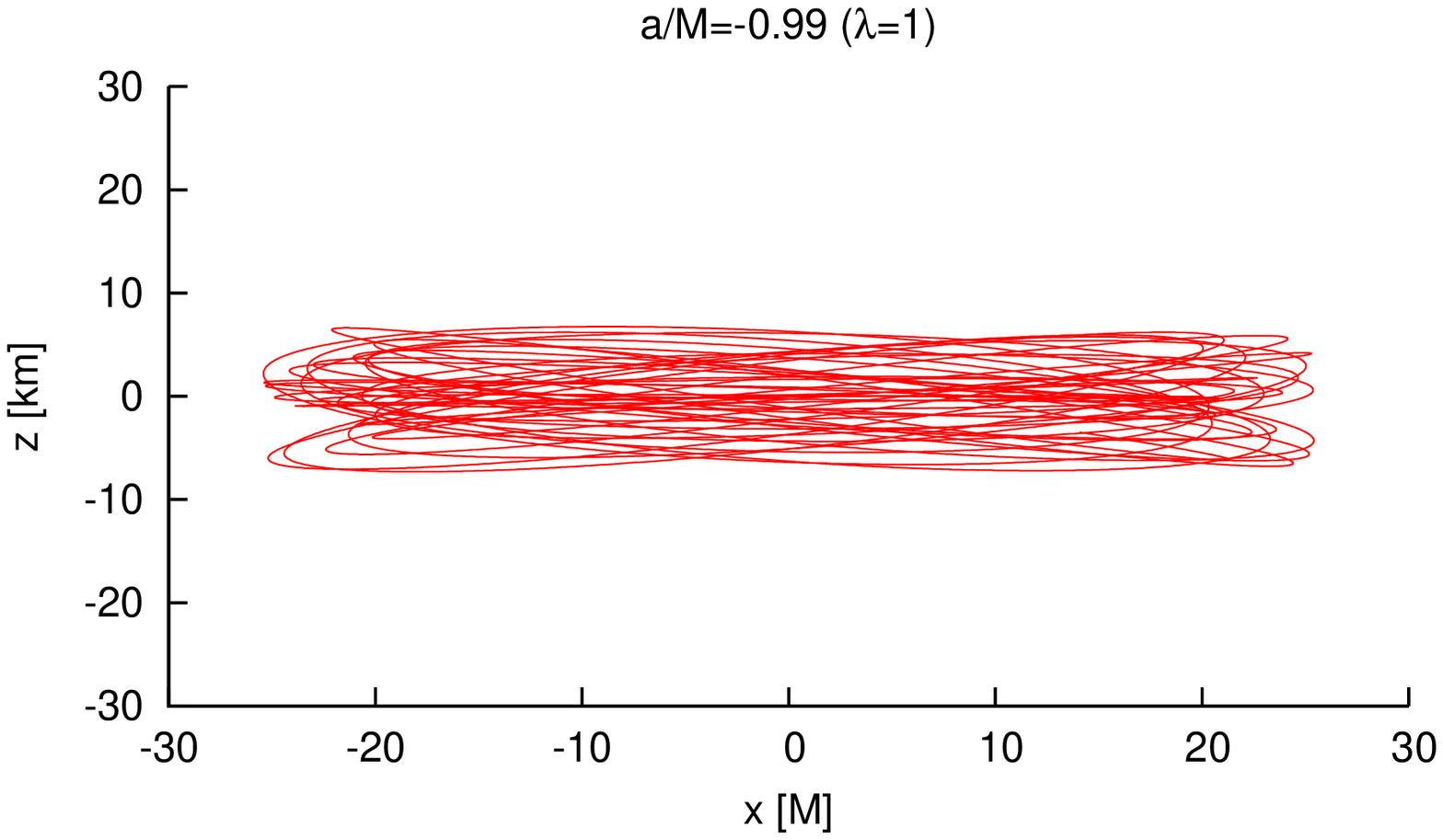}
  \includegraphics[width=0.325\textwidth]{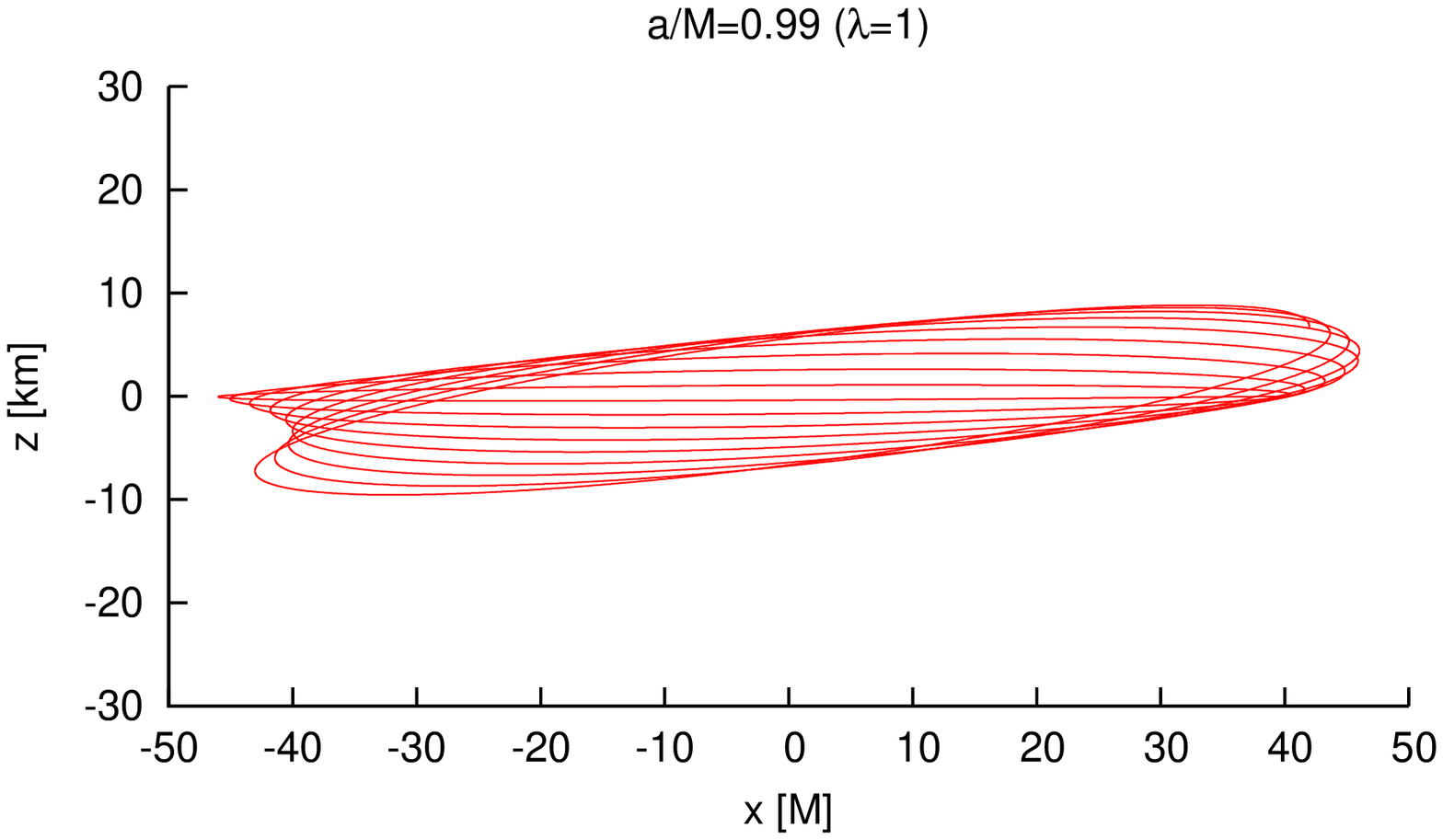}
\vspace*{-8em}
\caption{
  The projection of the orbit of a pulsar around a black hole onto the $x$-$z$ plane
      for $r=10M$, $r=20M$ and $r=40M$. The $z$ motion for separations in terms of $M$, with the scale in km, is independent of the black hole mass ($M=  2.0 \times 10^{6}{\rm M}_{\odot}$.
      $10^{5}{\rm M}_{\odot}$ or $10^{3}{\rm M}_{\odot}$).
  The black-hole spin parameter is $|a/M| = 0.99$,
       with prograde orbits in the top row and retrograde orbits in the bottom row.
  Other parameters are the same as those in Figure \ref{fig-5}.  Without spin-curvature coupling ($\lambda = 0$), the neutron star would not lift out of the $x$--$y$ orbital plane.
   }
\label{fig-6}
\end{center}
\end{figure}

\begin{figure}
\begin{center}
\vspace*{0.2cm}
  \includegraphics[width=0.325\textwidth]{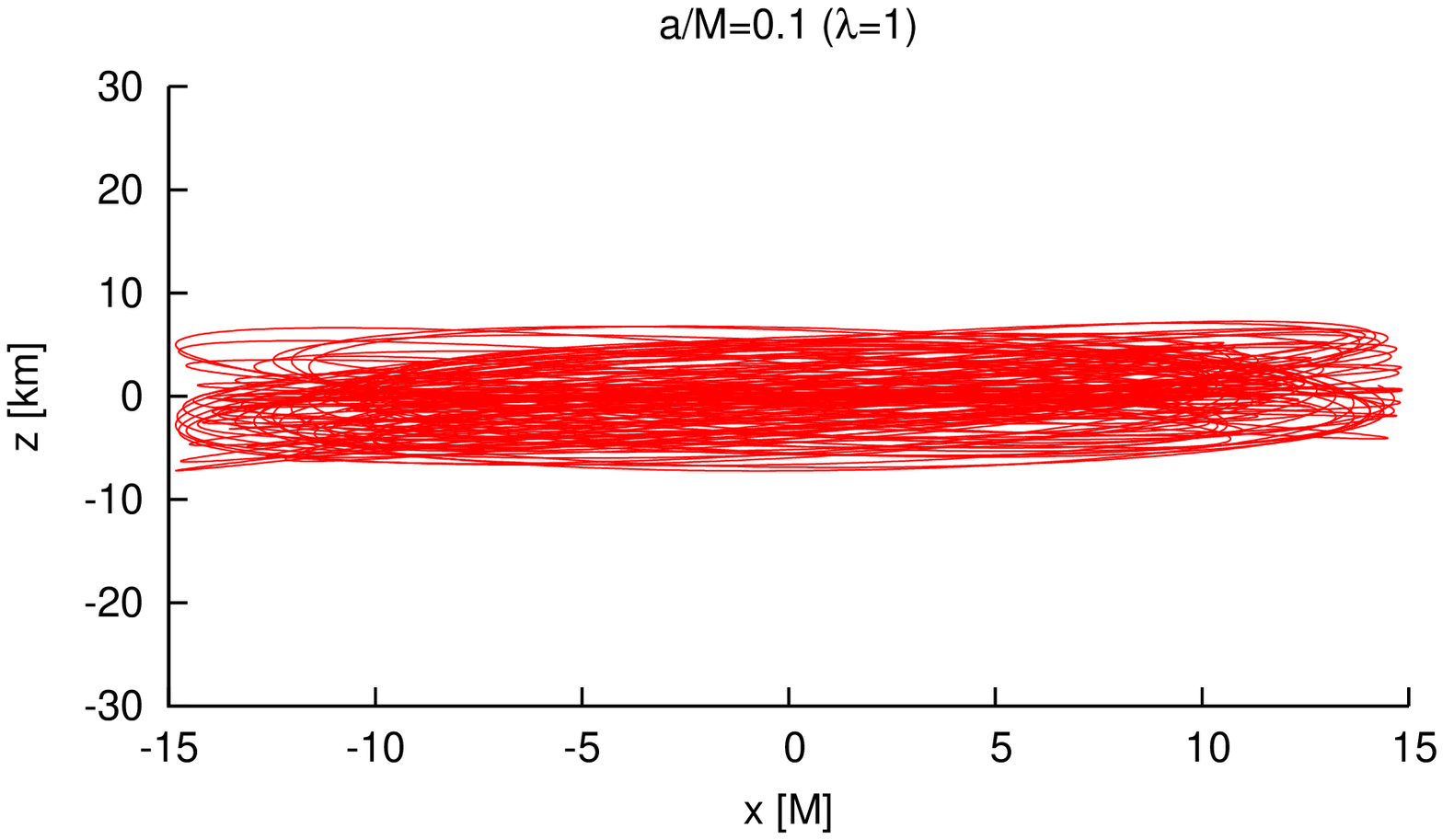}
  \includegraphics[width=0.325\textwidth]{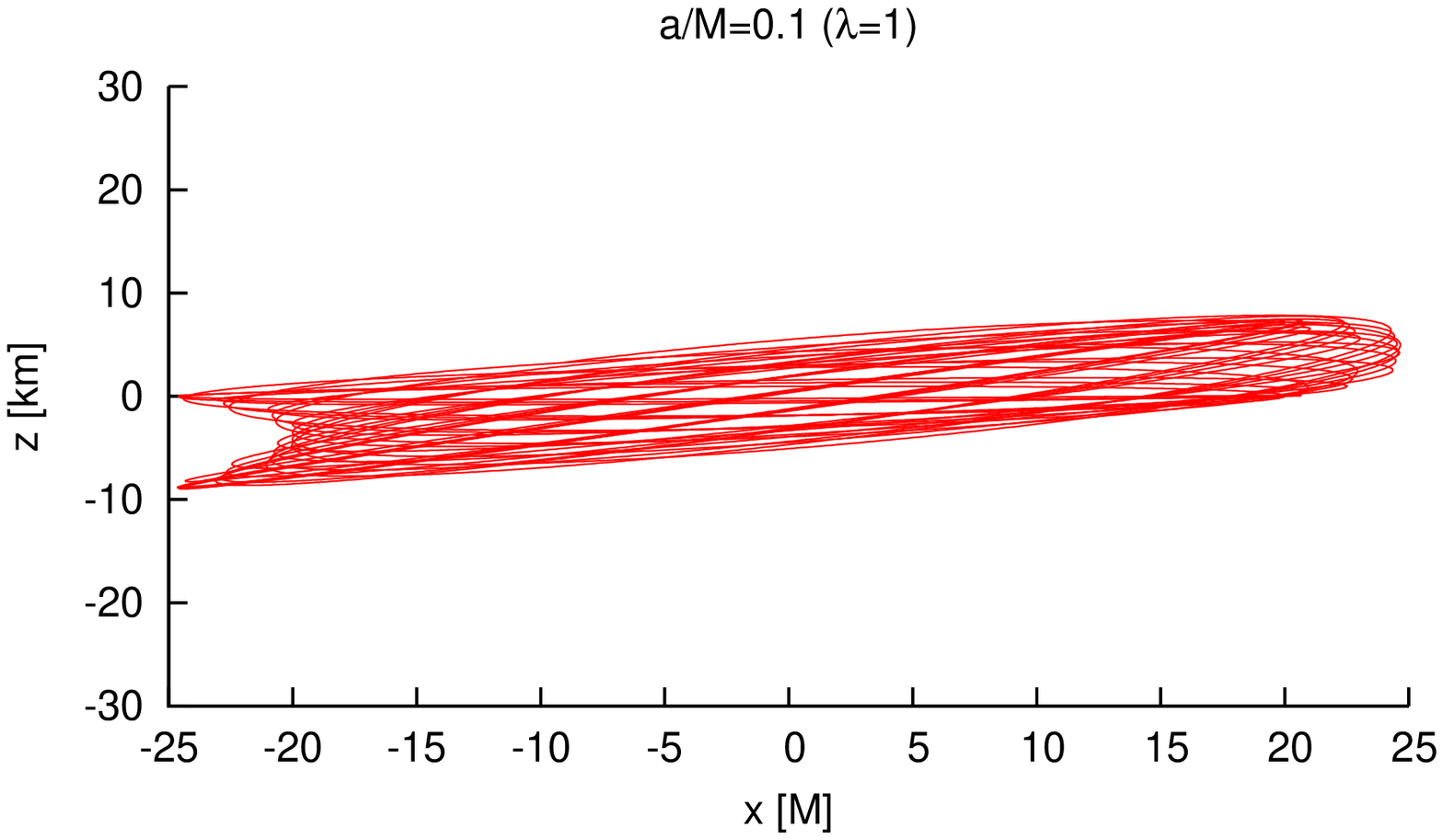}
  \includegraphics[width=0.325\textwidth]{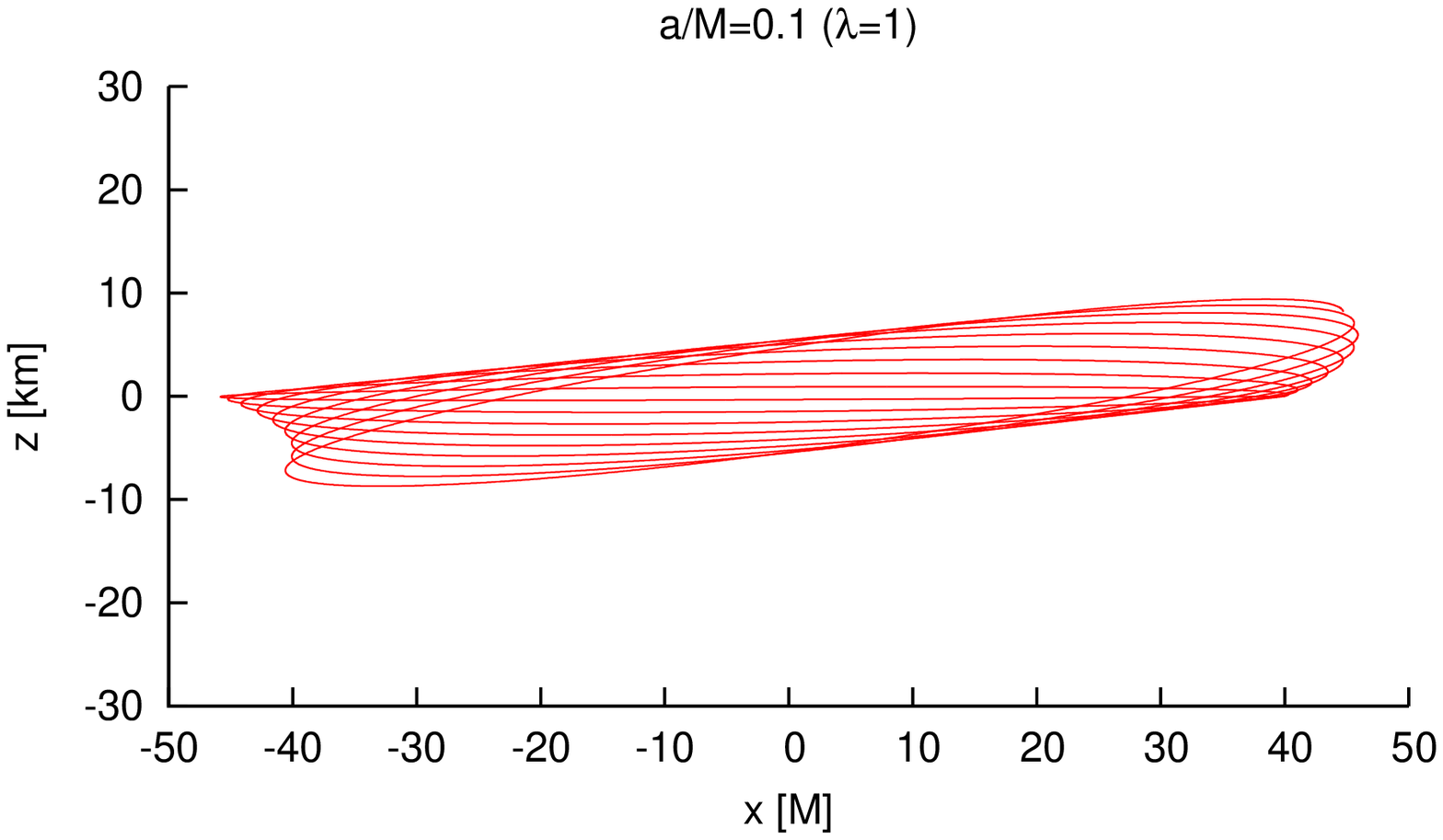}
\vspace*{-2.5cm}
\caption{
  The projection of the orbit of a pulsar around a slowly rotating black hole, black-hole spin parameter $a/M = +0.1$,  onto the $x$-$z$ plane for initial orbital separations of $r=10M$, $r=20M$ and $r=40M$. Vertical, $z$, motion amplitude is independent of central black hole mass.
      for $M=  2.0 \times 10^{6}{\rm M}_{\odot}$ (left panel)
      $10^{5}{\rm M}_{\odot}$ (middle panel) and $10^{3}{\rm M}_{\odot}$ (right panel).
The amplitude of the vertical motion is similar for all the cases computed, from lower mass to higher mass black holes, and from lower spin to higher spin black holes. For varying initial separations, the amplitude of all the cases is similar but the paths are different. 
   }
\label{fig-7}
\end{center}
\end{figure}

Figure \ref{fig-6} shows the out of plane motion in terms of physical units instead of in terms of mathematical $M$ units. It is clear that the $z$-range and the trajectory are independent of the $M/m$ ratio. With no spin-curvature coupling ($\lambda=0$) the motion of the pulsar is planar and does not leave the $x$--$y$ plane. However, in relative terms, the out of plane motion is smaller relative to the diameter of the orbit for more massive black holes. This means that the effects of spin-curvature coupling will become apparent in an observed pulsar signal at higher inclinations for lower mass black holes. The effect of the Kerr geometry on an emitted pulsar light signal from a neutron star at superior conjunction depends on the $M$ unit distance from the black hole as it passes above or below the black hole on its way to Earth. 
If a pulsar signal were interpreted without accounting for spin-curvature coupling ($\lambda = 0$) then a low mass black hole - pulsar pair would appear to have a higher inclination than the actual value while the difference for a million solar mass black hole - pulsar pair would be much smaller between the $\lambda=0$ and $\lambda=1$ models. The motion of the neutron star's spin axis direction is not appreciably affected by the spin-curvature coupling with the de Sitter precession and the Lense-Thirring effect being very much the same in both  the $\lambda=0$ and $\lambda=1$ models. The physical amplitude of the out of plane motion is, for our calculations, independent of the $a/M$ ratio as well (Figure \ref{fig-7}), although the exact path taken does depend on the black hole spin. This means that the amplitude of the out of plane motion is entirely a function of the pulsar spin rate which is an unexpected result.
    

\begin{figure}
\begin{center}
\vspace*{-2em}
  \includegraphics[width=0.325\textwidth]{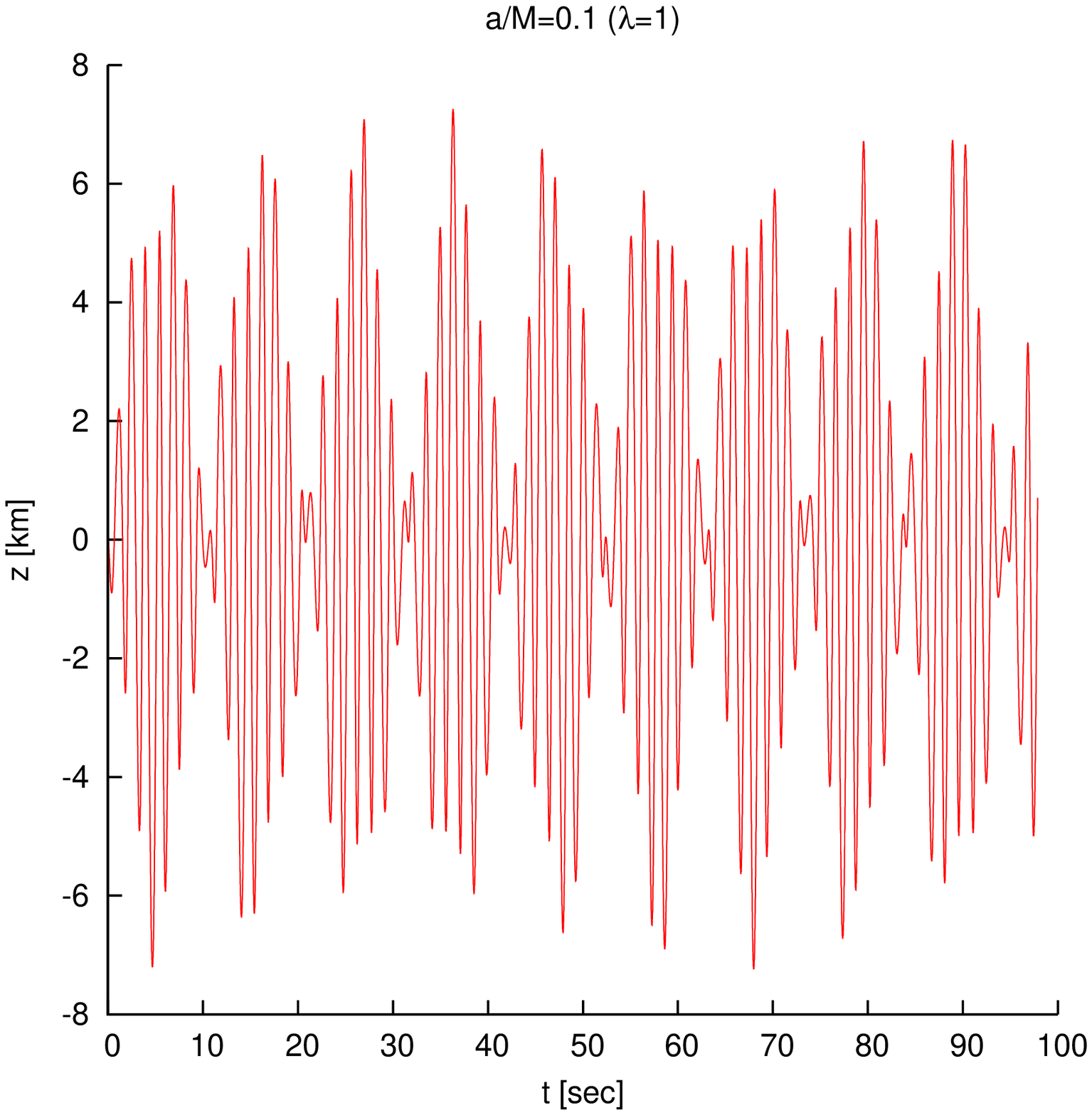}
  \includegraphics[width=0.325\textwidth]{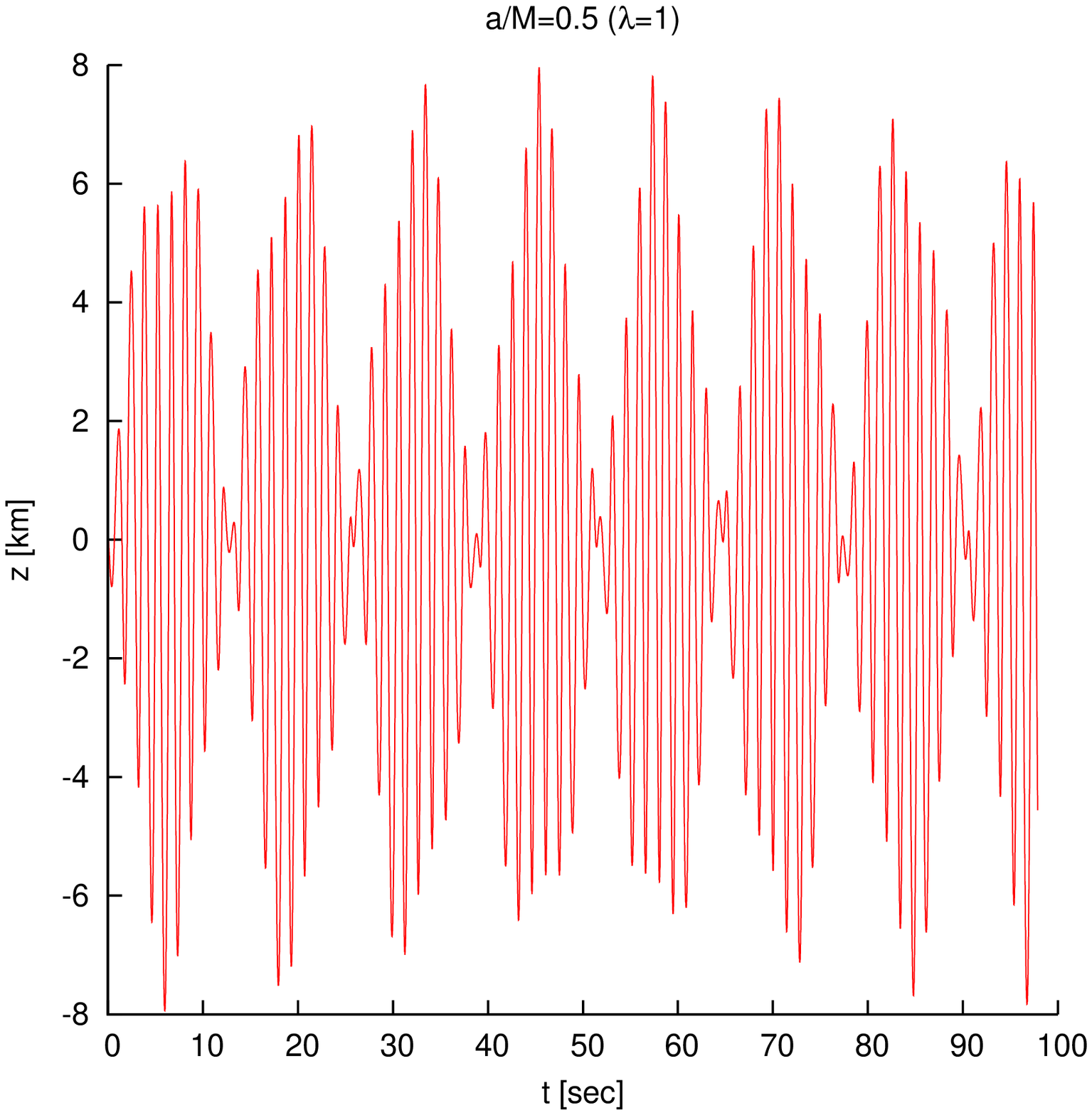}
  \includegraphics[width=0.325\textwidth]{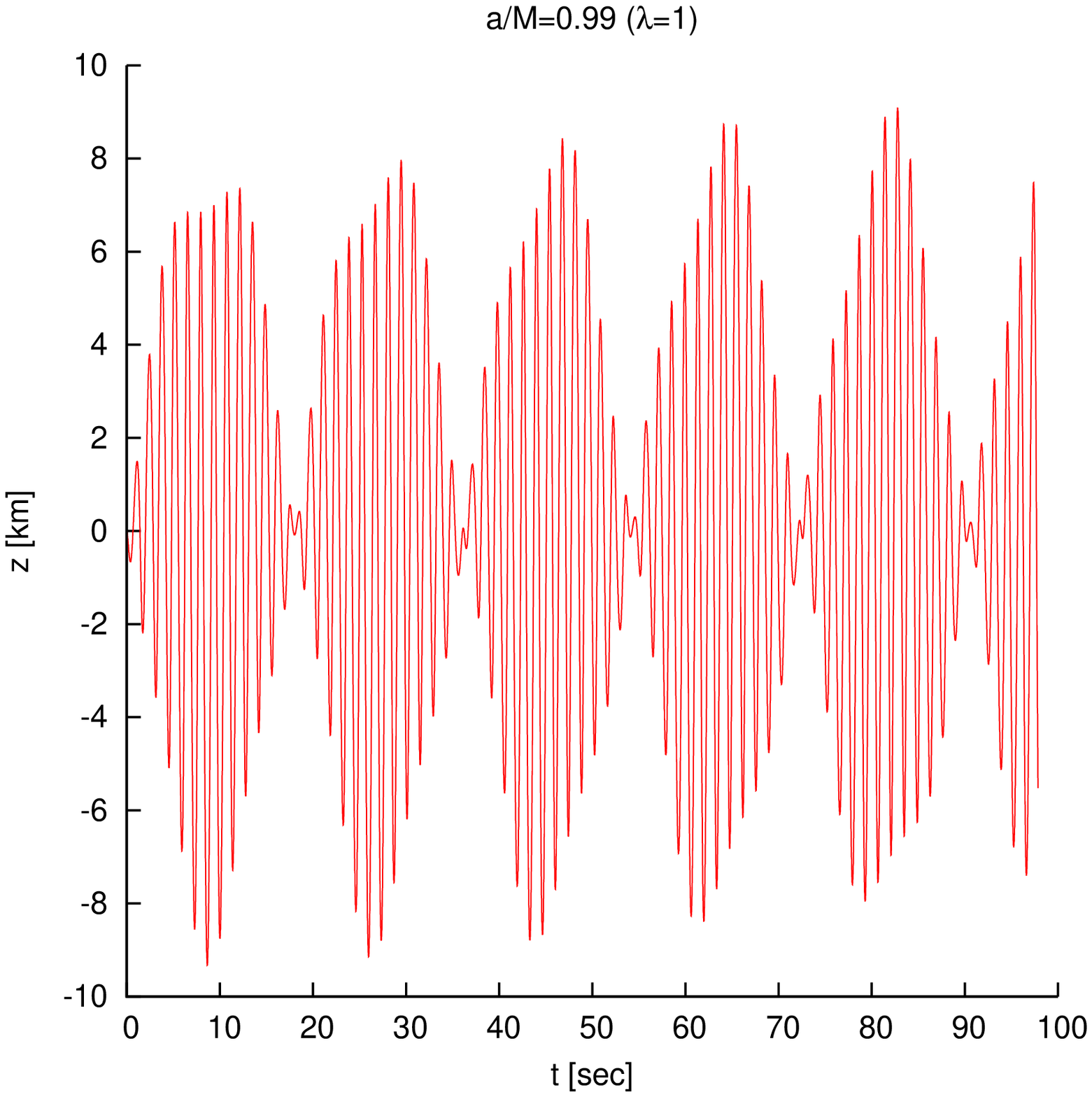}

  \vspace*{-6em}

  \includegraphics[width=0.325\textwidth]{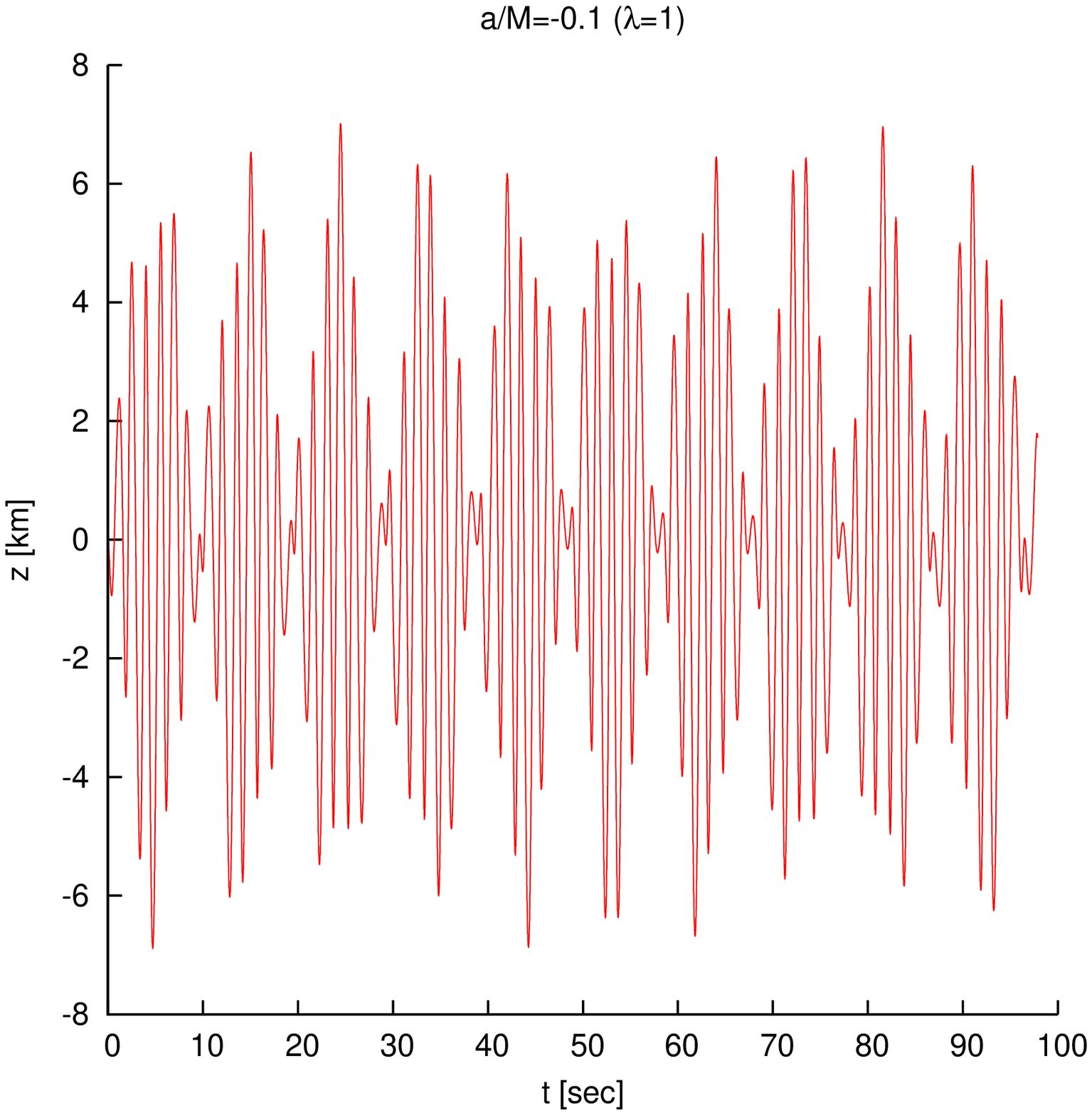}
  \includegraphics[width=0.325\textwidth]{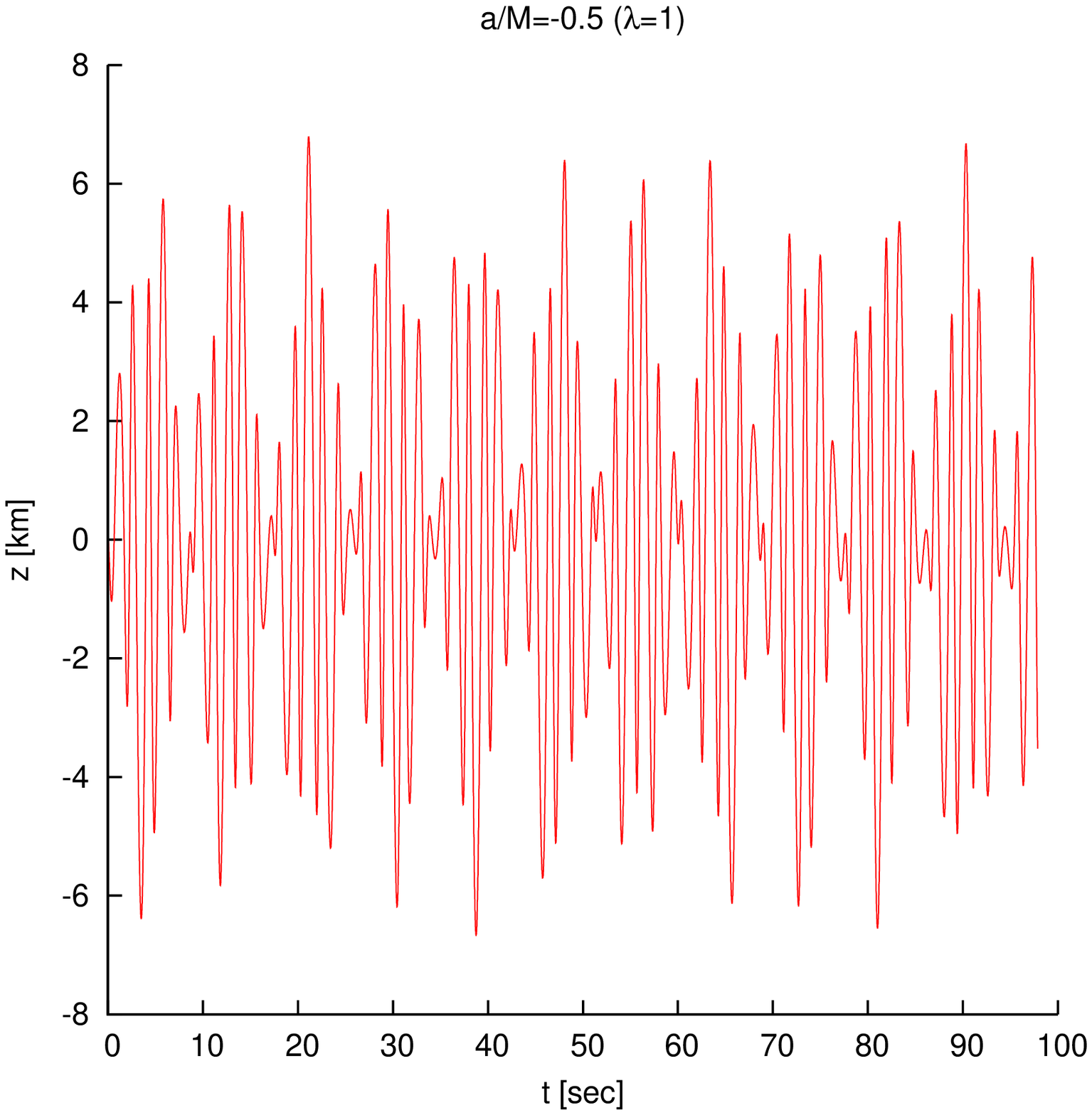}
  \includegraphics[width=0.325\textwidth]{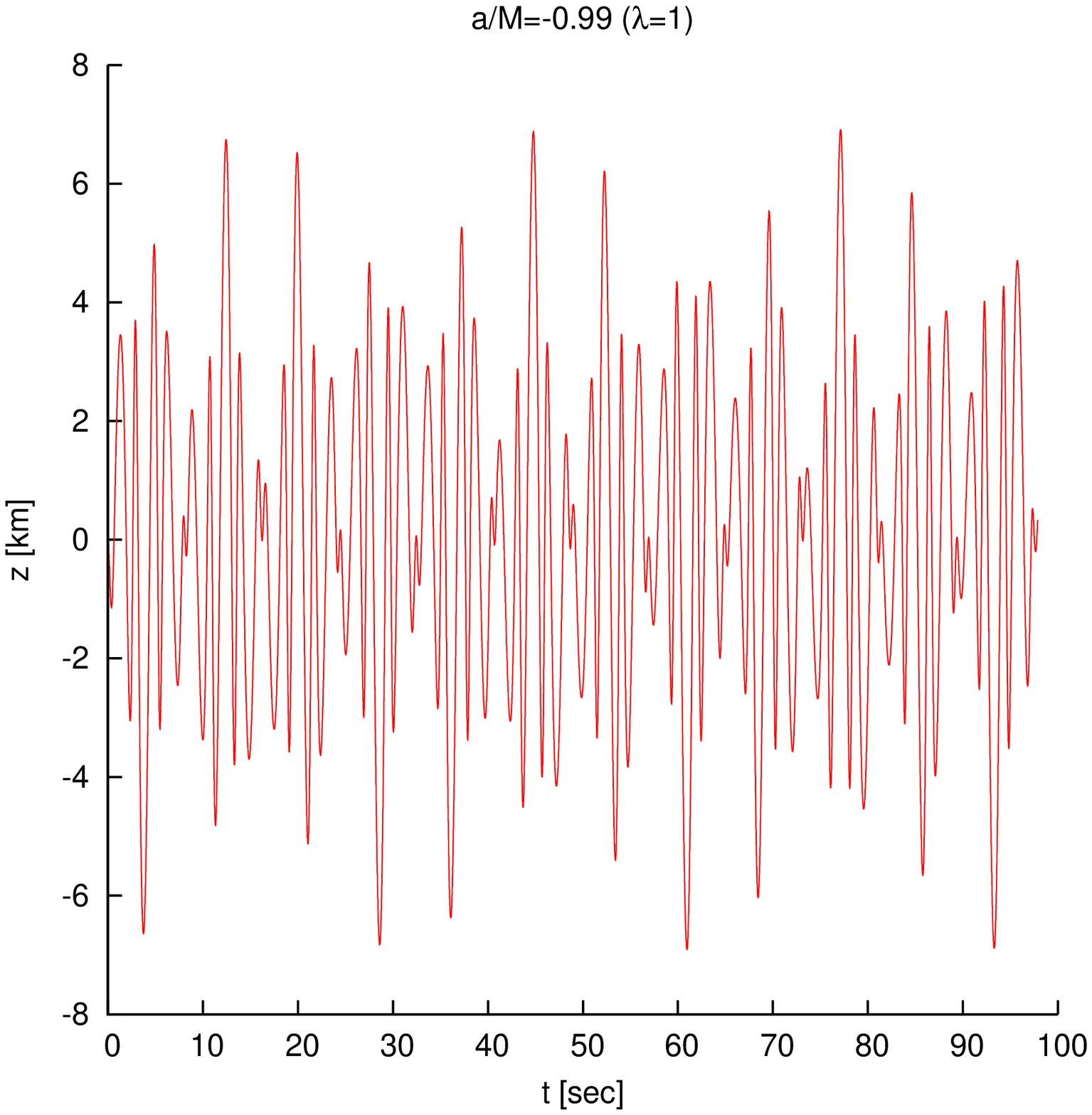}
\vspace*{-2em}
\caption{
  Time course of the $z$-component of the pulsar motion. The top rows show the prograde cases and the bottom rows show the retrograde cases for $|a/M|=0.1$, 0.5, 0.99 from left to right for an initial $r=10M$. The time courses are independent of the mass of the central black hole.  Varying the initial separation will vary the time course followed but the amplitude will be similar at $\sim 8$ km.
   }
\label{fig-8}
\end{center}
\end{figure}

The time course of the out of plane motion is shown in Figure \ref{fig-8} for three different values of $a/M$. 
Although the amplitude is similar, the time courses are different. Varying the initial separation will also result in different time courses with similar amplitudes for the ms pulsar simulated here. The $z$ signal is potentially observable through variation in pulse arrival time over arrival times that would result if the pulsar stayed in a planar orbit. Ray-tracing solutions and simulations of the pulsar signal are beyond the scope of this work but an order of magnitude effect may be estimated. 
For reasonable viewing orbital inclinations (say $i \sim 45^\circ$)
  the path length of the ray from the pulsar to the Earth will vary by $\sim \pm 5$ km from the path length of a ray from a pulsar in an otherwise similar planar orbit with a frequency that is roughly twice the pulsar's orbital frequency. 
Thus frequencies outside of those predicted by models of planar orbital motion will be introduced into the pulsar signal. The light ray path length change between planar and non-planar motion translates to a timing change on the order of $\pm 10 \mu$s. 
Such timing changes are readily detected in secular observations of pulsars, especially millisecond pulsars \citep{Lorimer08}. The fastest orbital periods, with their attendant Doppler shifts, of the models investigated here are about 2 sec (it is not constant) for the pulsar in a prograde orbit  at $r =10 M$ from a 10$^{3}$M$_{\odot}$ black hole and about 10 sec for $r = 40 M$. The other cases are less extreme with the 10$^{5}$M$_{\odot}$ black hole cases having periods of roughly 200 sec for $r = 10 M$ and 900 sec for $r = 40 M$; the $2 \times 10^{6}$M$_{\odot}$ case orbital periods range from roughly 4000 sec for $r = 10 M$ to 19000 sec for  $r = 40 M$. Our models predict similar timing changes for all cases. The orbital period of the Hulse-Taylor pulsar PSR~J1915+1606 is $\sim$27900 sec \citep{Weisberg10}, so observing millisecond pulsars in systems like those modelled here presents no new technological challenge for the higher mass black holes. Since pulsars emit at a wide range of frequencies, detection of the faster orbiting systems should not be deterred by their rapidly changing Doppler shifts, but the analysis of the data may be challenging.


Our calculations have demonstrated that potentially observable orbital dynamics of a ms-pulsar (a very fast spinning neutron star) around a massive black hole is affected by modifications to the space-time structure that define the black hole's gravity caused by the spin of the pulsar. The orbital motion of a ms-pulsar and the relative geometry as viewed from Earth depend on the spin of the black hole and on the mass ratio of the black hole and the neutron star. The potentially observable effects should be higher for high-inclination, low-mass black holes that have high spin rates and for pulsars in retrograde orbits around the black holes.

\section{Astrophysical implications}
\label{sec-imp}

\subsection{Millisecond pulsars as probes to the space-time around rotating black holes}

The use of pulsars to probe the space-time around black holes has been proposed by many workers \citep[e.g.][]{Wex99,Pfahl04,Wang09,Liu12}. Generally, the proposed diagnoses involve one of two approaches. The first considers the effects of the black hole's presence on the propagation of the pulsar's beamed emission, e.g.\ gravitational bending/lensing \citep[e.g.][]{Wang09,Nampalliwar12} and Shapiro time delay \citep[e.g.][]{Laguna97}. The second considers the spin precession (and nutation) of the pulsar induced by spin-orbit coupling or by the spin interaction with the black hole \citep[e.g.][]{Wex99,Liu12}. In both cases, the orbital motion of the pulsar follows a geodesic in the space-time determined by the black hole. For a spinning object orbiting around a rotating black hole along a geodesic, one can construct two Hamiltonians, corresponding to the spin-orbit coupling and to the spin-spin interaction with the black hole respectively, in the limits of slow motion and of a large mass ratio between the black hole and the spinning object \citep{Iorio12}. For a pulsar revolving around a massive galactic black hole, these spin-orbit and spin-spin Hamiltonians perturb the pulsar's Keplerian orbit and drive the pulsar's spin to precess. It can be shown that with a suitable choice for the alignment of the black-hole spin vector with respect to the azimuthal axis in the reference frame of the observer, the conventional Lense-Thirring precession of the Keplerian orbital elements of the pulsar can be derived from the two Hamiltonians \citep[see][]{Iorio12}. However, these Hamiltonians do not take account of the spin-curvature coupling between the pulsar and the black hole. Spin-curvature coupling occurs physically because the spin of the pulsar will modify the Kerr metric of the space-time. In the presence of spin-curvature coupling, the pulsar's motion does not follow a Kerr space-time geodesic and this motion is modeled by the MPD equation (\ref{MPD-x1}) without explicitly modifying the Kerr metric. Using the MPD equations to model the ms-pulsar motion around a rotating black hole gives proper consideration to the spin-curvature coupling. The results in \S \ref{sec-orb} are general, and they recover the results derived from conventional treatments of spin-orbit coupling and spin-spin interaction between the pulsar and the black hole, by taking appropriate limits for the parameters. For instance, the spin precession of the pulsar due to parallel transport along the geodesic can be obtained from the MPD equations (\ref{MPD-x1}), (\ref{MPD-x2}) and (\ref{MPD-x3}) in the limits of $\lambda \rightarrow 0$ and $M/m \rightarrow \infty$.

Pulsar timing is potentially a powerful tool for probing the space-time around black holes,
  especially in the strong gravity regime \citep[see e.g.\ the review by][]{Cordes04}.
As indicated in the studies of \cite{Wex99} and \cite{Liu12},
  measuring the pulsar-spin precessions can determine the rotation rates of the massive central black holes in galaxies,
  such as that in the Galactic Centre, with good accuracy,
Thus, pulsar timing provides an alternative to the current methods of black-hole spin determination, such as X-ray line spectroscopy.
In \S \ref{sec-orb} we have shown that
  the spin interaction between the pulsar and the black hole
  can also cause substantial variations in the pulsar's orbit (see Figures \ref{fig-5} to \ref{fig-7})
  in addition to the well-understood pulsar spin precession (and nutation).
These variations are non-negligible,
  and they will modify the arrival time of the pulsar's emission pulses.
Our calculations show that the complexity and the relative amplitude of the orbital variations increase with the black hole's rotation
  when other parameters are kept constant.
The variations are dramatic when the pulsar is in a retrograde orbit (see e.g.~Figure \ref{fig-5}).
At certain orbital separations,
  complex orbital motions occur for a wide mass range covering that of the predicted intermediate-mass black holes in globular clusters
  and that of the central massive black holes in galaxies.

The spin-curvature coupling between the pulsar and the black hole causes the pulsar to deviate from Kerr geodesic motion.
The pulsar orbits show large-amplitude complex orbital variations, which are easily distinguishable for pulsars orbiting around low-mass black holes because of the larger ratio of the amplitudes of the out-of-plane to in-plane motion.
Since pulsars have a very narrow mass range around 1.5M$_{\odot}$ \citep[][]{Lorimer08, Steiner10, Lattimer11},
  knowing the mass ratio between the black hole and the pulsar is essentially the same as knowing the black hole mass.
Thus, analyses of pulse arrival time modulations caused by the orbital variation and the precession of the ms-pulsar spin
  will give us very accurate measurements of the mass as well as the rotation rate of the black hole that the pulsar is orbiting.

\subsection{Pulsars around central black holes in spheroid systems}

It is believed that a large population of stellar remnants reside in a small parsec-scale region around Sgr A$^{*}$,
 the compact radio source at the Galactic Centre.
On one hand, studies \citep[e.g.][]{Freitag06} have shown that there could be as many as $10^{3}$ neutron stars within a parsec from the Galactic Centre.  One the other hand the density profile of stars near Sgr A$^{*}$ is different from the distribution expected for a dynamically relaxed distribution of stars near a $10^{6}$M$_{\odot}$ black hole \citep{Bartko10} leading to the prediction of a somewhat lower number of neutron stars.
Some of the these neutron stars would be in binaries, thus they would have been spun up by accretion \citep{Alpar82} to become ms-pulsars.
Recent observations include the {\em Swift} discovery of a soft gamma repeater (SGR likely a magnetar) within $\sim$0.1 parsec of Sgr A$^{*}$ \citep{Kennea13,Mori13}. A magnetar plus an undetected pulsar population might indicate lower numbers of ordinary pulsars \citep{Dexter13}, however others argue that such a conclusion is premature and argue for a population of $\sim 10^{3}$ neutron stars in the central parsec of the Galaxy \citep{Chennamangalam13}.
The number estimate of $\sim 10^{3}$ neutron stars in the central parsec of the Galactic Centre is based on models
  assuming the presence of $20,000-40,000$ stellar-mass black holes in the same region \citep[see][]{Miralda00}.
Without this cluster of black holes, the central concentration of neutron stars could be significantly higher \citep{Freitag06}.
Following this line of reasoning, we would expect that galaxies with spheroids similar to that of our Galaxy would have about 1000 neutron stars around their central black holes.
Large elliptical galaxies would have larger central neutron-star populations while
  dwarf spheroidal galaxies and small elliptical galaxies would have smaller populations accordingly.
Some of the neutron stars will inevitably fall into the central black hole in some galaxies
  and it is possible that such galaxies will have pulsars in close orbits around their central black holes.
It is also possible that some pulsars are actually in close orbits around the central black hole in our Galaxy,
  although detecting them is a great technical challenge currently \citep[see][]{Bates11}. If, however, the lower mass stars have a shallow density profile as observed in the Galactic center, the absence of a BH cusp would not necessarily imply a higher density of neutron stars \citep{Antonini2012}, so the distribution of neutron stars in galactic centres should be considered very uncertain.

Globular clusters are also known to contain a large number of neutron stars.
Because of mass segregation the majority of the neutron stars have sunken to the cluster cores.
There is evidence that the most massive globular clusters in our Galaxy contain more than $\sim 1000$ neutron stars.
The retention of a large population of neutron stars in globular clusters is usually explained by
  models in which the progenitor stars of these neutron stars were in binary systems that retained the neutron stars in spite of their supernova kicks
  \citep{Drukier96,Pfahl02}.
Millisecond-pulsars are believed to be remnant descendants of binary systems \citep{Alpar82}.
A substantial fraction of pulsars in the globular clusters are in fact ms-pulsars \citep{Camilo05},
  and 30 pulsars with spin period shorter than 10~ms have been found in the globular cluster Terzan 5 alone
  \citep{Ransom05,Ferraro11}.
It has long been proposed that globular clusters could contain central black holes with substantial masses
  \citep{Bachall75, Silk75}.
Extrapolation of the the $M$-$\sigma$ relation predicts that the central black holes in globular clusters
  would have masses $M_{\rm bh} \sim 10^{3}-10^{4}{\rm M}_{\odot}$ \citep[see][]{Gebhardt02}.
There are observational claims that there are central black holes in globular clusters \citep[e.g.][]{Newell76, Gerssen02, Maccarone04},
  but there are also studies providing alternative explanations \cite[e.g.][]{Illingworth77, Baumgardt03, Kirsten12}.
The definitive search for intermediate-mass black holes in globular clusters is still ongoing.

The high number density of neutron stars in globular cluster cores along
  with a large fraction of those neutron stars being ms-pulsars
  implies that  there is a good chance that a ms-pulsar is revolving in a close orbit around an intermediate-mass black hole
  if there is, in fact, a single intermediate-mass black hole present in globular cluster cores.
Moreover, there is roughly a 50/50 chance that the pulsar is in a retrograde orbit
 since the stars in globular clusters do not have strong preference for direction of rotation.
As shown in our calculations,
  ms-pulsars orbiting around $10^{3}$M$_{\odot}$ black holes should have very distinguishable dynamical signatures.
From these signatures we can infer the mass ratio of the two objects, and hence the black hole mass accurately, as well as the spin of the black hole.

Conventional methods for determining the masses of black holes in spheroidal systems,
  such as stellar kinematics, are not effective for more massive black holes.
In contrast, pulse timing analyses of spin-interactions between pulsars and black holes are effective for black holes below $\sim 10^{6}$M$_{\odot}$,
  which is complimentary to the stellar kinematic methods.
The Square Kilometer Array Telescope, to be in operation in the near future,
  will discover about 20,000 pulsars in the Galaxy of which 6,000 will be ms-pulsars \citep{Smits09}.
It will also allow a systematic search for ms-pulsars which are beyond our Galaxy and the two Magellanic Clouds.
This sensitivity opens up the opportunity to use pulsar timing to measure the masses of central black holes in the Local Group galaxies and other nearby galaxies,
  and in their globular clusters.
  Such pulsar timing would thus settle the disputes regarding the existence of intermediate-mass black holes
  and properly establish the low-end of the $M$-$\sigma$ relation for central black holes and their host spheroids.
The pulsar timing of spin precession, as shown by other workers, e.g. \cite{Wex99} and \cite{Liu12}, and of spin-curvature induced orbital variations,
  as shown in this work, will also provide accurate measurement of the spins of those black holes.

The MPD equations (\ref{eqn-mMPD1})--(\ref{eqn-mMPD3}) are an approximation of the general relativistic dynamics
  that would occur between a spinning black hole and a spinning neutron star.
In a complete, less tractable, treatment the space-time metric of the system would be
   a non-linear combination of the Kerr metric of the spinning black hole
   and the Kerr metric of the spinning neutron star and the neutron star would move along a geodesic in such a metric.
A more complete treatment would also model gravitational radiation
   so we need to be assured that the gravitational wave time scale is much longer than the dynamical scale of the orbital motion
   in order for the MPD model to be a good approximation.
The time scale for the change in the orbital period $P_{\rm orb}$ due to gravitational radiation is
\begin{eqnarray}
\tau_{\rm gw} &  \sim &  
  \frac{5\  a_{\rm orb}^{4}}{96  \ mM(m+M)} f(e)^{-1}
\end{eqnarray}
\citep[see e.g.][]{Misner73,Fang83},
  where $a_{\rm orb}$ is the orbital separation,  
  and $f(e)$ is a function of the orbital eccentricity, which is given by 
\begin{eqnarray} 
 f(e) & = &  (1-e^2)^{-7/2} \left[ 1+\frac{73}{24}e^2  + \frac{37}{96}e^4 \right]  \ . 
\end{eqnarray} 
Setting $a_{\rm orb} = \varpi M$, we have 
\begin{eqnarray}  
  \frac{\tau_{\rm gw}}{P_{\rm orb}} & \sim  &  \frac{5}{192 \pi} \frac{\varpi^{5/2}}{f(e)} \left( \frac{M}{m} \right) \left( \frac{M}{m+M} \right)^{1/2} \ .   
\end{eqnarray} 
In this work we investigated systems with $10^{3}{\rm M}_{\odot} \leq M \leq 2 \times 10^{6}{\rm M}_{\odot}$ and $10 M \leq a_{\rm orb} \leq 40 M$,
   so $\varpi \sim (10 \mbox{ -- } 40)$ and $M/m > 6 \times 10^2$.  
Moreover, $f(e) \sim 1$.  
   Hence, $\tau_{\rm gw}/P_{\rm orb} \sim (10^3 \mbox{ -- } 10^8) \gg 1$, 
   justifying our employment of the MPD approximation. 
While gravitational radiation loss is not substantial in single pulsar timing observations,
   longer term observations must take into account the gravitational radiation effects, such as changes in the orbital period.

The MPD equations provide a simple way of modelling the significant out of plane motion of a pulsar orbiting a massive black hole.
Pulsars are also known to occur in pairs or as neutron star binaries in which one of the neutron stars is a pulsar
\citep[e.g.\ the famous PSR~B1913$+$16,][]{Hulse75}.
So there is the possibility that similar neutron star binaries also orbit massive black holes.
The motion of neutron star binaries has been analyzed recently \citep{Remmen13} in the rigid mass ring current approximation. 
However, under circumstances similar to those modelled here, 
  substantial out of plane motion of the binary neutron star would occur as the pair orbited the black hole.




\section{Conclusion}

Signals from pulsars orbiting in the strong field of moderate to massive black holes offer a means to determine the mass and spin of the central black hole.
The non-linear nature of the general relativistic field equations $G_{\mu \nu} = 8 \pi T_{\mu \nu}$
   means that the computation of the motion of anything more than a test particle in the strong field near a black hole generally requires numerical methods.
In particular, the mass and spin of an orbiting neutron star will change the space-time geometry $G_{\mu \nu}$ through its stress-energy tensor $T_{\mu \nu}$.
Until now, analysis of the motion of a pulsar near a black hole and that motion's effect on the observed pulsar signal have considered the motion of the pulsar
   as a test particle moving along a geodesic in the Kerr space-time of a rotating black hole.
The effect of the mass and spin in the pulsar's stress-energy tensor on the pulsar's motion had not been previously considered.
Here we have demonstrate that effect through the approximation given by the MPD equations.

The MPD equations used here consider the effect of the first two moments of the pulsar's stress-energy tensor on the pulsar's motion.
Our computations for the astrophysically important cases corresponding to intermediate mass black holes and the nuclear black holes of low-mass galactic spheroids
  show that the pulsar's spin leads to significant motion out of the usual orbital plane.
The extent of the out of plane motion of a $1.5 {\rm M}_{\odot}$ pulsar becomes comparable to the extent of the orbit's radius
  for black holes of masses $10^{3}{\rm M}_{\odot}$.
This motion therefore needs to be accounted for to properly interpret the timing of pulsar signals from a pulsar that is closely orbiting any intermediate mass black hole
  that may exist in globular cluster cores.
Models of observed pulsar timing that use the MPD equations will therefore provide accurate measurements
   of masses and spins of central black holes in globular clusters and nuclear black holes in the galactic spheroids at the low end of the $M$-$\sigma$ relation.


\section*{Acknowledgments}
We thank Roberto Soria for discussions on the mass spectrum of astrophysical black holes.
KW's visit to University of Saskatchewan was supported by the University of Saskatchewan's
  Role Model Speaker and Visiting Lecturer funds.

\label{lastpage}

\end{document}